\documentclass[conference]{IEEEtran} 
\usepackage{amsmath,amsfonts}
\usepackage{algorithmic}
\usepackage{algorithm}
\usepackage{array}
\usepackage[caption=false,font=normalsize,labelfont=sf,textfont=sf]{subfig}
\usepackage{textcomp}
\usepackage{stfloats}
\usepackage{url}
\usepackage{verbatim}
\usepackage{graphicx}
\usepackage{cite}
\usepackage{hyperref}
\begin{document}

\title{Efficient and Privacy-Protecting Background Removal for 2D Video Streaming \\ using iPhone 15 Pro Max LiDAR}

\author{
    \IEEEauthorblockN{Jessica Kinnevan\IEEEauthorrefmark{1}, Naifa Alqahtani\IEEEauthorrefmark{2}, and Toral Chauhan\IEEEauthorrefmark{3}}\\
    \IEEEauthorblockA{\IEEEauthorrefmark{1}Department of Computer Science, Iowa State University, Ames, IA, USA} \\
    \IEEEauthorblockA{\IEEEauthorrefmark{2}Department of Computer Science and School of Education, Iowa State University, Ames, IA, USA} \\
    \IEEEauthorblockA{\IEEEauthorrefmark{3}Department of Computer Science, Iowa State University, Ames, IA, USA} \\
    \thanks{The authors would like to thank Professor Alexander Stoytchev of Iowa State University for his guidance and support.}
}
% The paper headers
%\markboth{Journal of \LaTeX\ Class Files,~Vol.~14, No.~8, August~2021}%
%{Shell \MakeLowercase{\textit{et al.}}: A Sample Article Using IEEEtran.cls for IEEE Journals}

% \IEEEpubid{0000--0000/00\$00.00~\copyright~2021 IEEE}
% Remember, if you use this you must call \IEEEpubidadjcol in the second
% column for its text to clear the IEEEpubid mark.

\maketitle

\begin{abstract}
Light Detection and Ranging (LiDAR) technology in consumer-grade mobile devices can be used as a replacement for traditional background removal and compositing techniques. Unlike approaches such as chroma keying and trained AI models, LiDAR's depth information is independent of subject lighting, and performs equally well in low-light and well-lit environments. We integrate the LiDAR and color cameras on the iPhone 15 Pro Max with GPU-based image processing. We use Apple's SwiftUI and Swift frameworks for user interface and backend development, and Metal Shader Language (MSL) for realtime image enhancement at the standard iPhone streaming frame rate of 60 frames per second. The only meaningful limitations of the technology are the streaming bandwidth of the depth data, which currently reduces the depth map resolution to 320x240, and any pre-existing limitations of the LiDAR IR laser to reflect accurate depth from some materials. If the LiDAR resolution on a mobile device like the iPhone can be improved to match the color image resolution, LiDAR could feasibly become the preeminent method of background removal for video applications and photography.

\end{abstract}


\section{Introduction}
\IEEEPARstart{I}{n} recent years, the rapid evolution of technology in consumer electronics like smart phones has significantly enhanced the capability to capture high-quality video content. Among these advancements, the integration of Light Detection and Ranging (LiDAR) sensors in these devices has opened new avenues for creative and technical applications in imaging. LiDAR sensors provide depth data for captured frames, enabling a more nuanced understanding of the spatial arrangement of elements within the scene.  

LiDAR is a transformative technology providing accurate data and extensive coverage \cite{zhang2024}. The depth sensing of LiDAR provides a unique feature of differentiating the foreground from the background elements when it comes to different and complex environments with low illumination levels or complex contrast.

LiDAR's proven capability in providing high-resolution, three-dimensional data presents a unique opportunity to enhance efficiency, accuracy, and decision-making processes in fields such as autonomous transportation, urban planning, environmental management, archaeology, agriculture, geology, and atmospheric research.

LiDAR as a technology has existed since the early 1960s. It is only recently, however, that LiDAR has been integrated into consumer-grade products available to the mass market. On October 23, 2020, Apple Inc. released the iPhone 12 Pro, the first consumer mobile device to incorporate LiDAR technology into its camera system. Its intended use was primarily for augmented reality and 3D scanning applications. Some applications, however, have begun to leverage LiDAR for 2D video and photo enhancements like background removal as a superior alternative to chroma keying or trained AI models, because it does not rely on having a well-lit subject or a high level of contrast between the subject and background.

This study aims to take 2D image processing with LiDAR on mobile a step further by using LiDAR to reduce image processing overhead in low-light conditions by focusing on exposure optimization for a given subject without processing their environment. The use of LiDAR will remove the need for computationally expensive 2D algorithms to isolate the subject in poor contrast scenarios so that more image enhancement can be done within each video frame while maintaining the standard video frame rate of 60 frames per second (fps).

\section{Related Work}
Before the integration of LiDAR into the iPhone Pro, consumer grade light field photography cameras developed by Lytro were available to the mass market as of early 2012 \cite{lytro2011}. The core concept of light field cameras is to employ a microlens array to sacrifice spatial resolution in favor of angular resolution, enabling the capture of images from multiple perspectives. Despite this innovation, light field cameras face significant measurement challenges due to their unique design. Primarily, the compromise on spatial resolution to facilitate depth estimation reduces measurement accuracy, which is a critical barrier to the further development and widespread application of light field cameras \cite{hu2023}. The consumer-grade Lytro cameras are no longer sold, and the technology is currently only available for machine vision applications through the German company Raytrix.

Samsung previously incorporated rear-facing LiDAR technology in several of its smartphone models including the Galaxy S10 5G, S20+, S20, and SOCELL Vizion 33D. This technology aimed to enhance camera focus and support augmented reality (AR) applications, similar to its use in Apple devices. In the Galaxy S20, LiDAR was employed in a ``Time-of-Flight" camera to scan objects and measure distances, though it was a simpler and lower quality version compared to Apple's LiDAR. After launching the Galaxy S21, Samsung discontinued the use of LiDAR sensors in its phones and has since ceased including the technology in all subsequent smartphone models \cite{fenstermaker2024}.

The TrueDepth technology integrated into the front facing camera of iPhone models X and later can also perform 3D scans. The technology is similar to LiDAR, though LiDAR technology has a longer range compared to Apple's TrueDepth camera, which powers Face ID using an array of infrared lasers effective only up to a few feet. In contrast, the LiDAR sensors used in the iPad Pro and iPhone Pro can operate effectively at distances up to five meters \cite{magnopus2020}. Face ID is most effective when the device is held at an arm's length, approximately 25-50 cm away from the face. The TrueDepth camera system is primarily used for 3D face authentication and recognition, while the LiDAR camera enhances augmented reality (AR) experiences by facilitating faster plane detection, thereby enabling new AR features \cite{vogt2021}.

No other consumer-grade cameras beyond the iPhone Pro 12 and later currently offer LiDAR or depth sensing technology, so there have not yet been many studies done leveraging this phone-based LiDAR system as it applies to 2D image processing. However, one recent study from January of 2024 investigated how LiDAR depth maps are utilized to improve the compression of images captured by the corresponding RGB camera, enhancing image quality and efficiency \cite{gnutti2024}. This study focuses primarily on compression and space reduction.

Our study contributes to this field by providing a robust and functional intermediate processing step to leverage both image and depth data to improve the process of background removal and replacement, with a special focus on optimizing performance in low-light conditions. These scenarios often present challenges for traditional image processing techniques due to the limitations of shape and color differentiation. By developing a GPU-based solution that integrates depth sensing with image processing, this study seeks to enhance the quality of resulting composite images in real time, thereby offering new capabilities for content creation, augmented reality, and beyond.  

\section{Experimental Approach}

\subsection{Equipment Used}
\begin{itemize}
\item 2 iPhone Pro Max 15 phones
\item 3 Macbooks with Apple Silicon processors
\item URCERI MT-912 light meter
\item Blackmagic Cinema Camera 2.5K (BMCC) with IR cut filter removed
\item Samwa LP10 Laser Power Meter
\end{itemize}
The BMCC was used only for visualizing the LiDAR laser pattern. The LP10 Laser Power Meter was used for measuring laser power levels for the purposes of background knowledge only (see \textit{D. Laser Safety Check}).
\subsection{Hardware Components}
\subsubsection{iPhone 15 Pro} Equipped with integrated color and LiDAR cameras, the iPhone 15 Pro Max captures depth information that can be synchronized with a companion color image. Fig. \ref{lidar_res_table} lists the relevant technical specifications of these two cameras.\vspace{0.1cm}
 
\subsubsection{MacBooks} All software development was performed on MacBooks with Apple silicon processors. They were selected for their superior computational capabilities for algorithm prototyping outside the iOS environment and for their compatibility with Xcode, the only IDE option for developing iOS applications.

\subsection{Software Components}

\subsubsection{SwiftUI and Swift} Swift is a programming language developed by Apple to simplify the development of applications for Apple devices. SwiftUI is a framework developed and maintained by Apple, employed in this study to design an intuitive and responsive user interface. SwiftUI allows developers to use declarative syntax to create UI elements, which can dramatically simplify the coding process and reduce development time. SwiftUI is particularly well-suited for rapid prototyping and dynamic interface design across all Apple platforms. Apple also provides a variety of well supported Swift-based frameworks for video and LiDAR related tasks, which made Swift essential when configuring a custom GPU render pipeline for this study.\vspace{0.1cm}

\subsubsection{Metal Shader Language (MSL)} Custom shaders written in MSL optimize the processing of depth data and image manipulation tasks. These shaders exploit the GPU's specialized capabilities, thus achieving peak performance in the handling of depth information and image data.
The Metal Shader Language, an extension of C++14, plays a pivotal role in the processing of LiDAR-generated depth data within this study's scope. Utilizing custom-written shaders, MSL is integral for achieving optimal image manipulation and data processing tasks. This language, specifically designed to harness the capabilities of GPU hardware, allows for the execution of high-performance parallel computing tasks. This feature is particularly beneficial for the analysis and processing of in-depth spatial data captured by the LiDAR scanner of the iPhone 15 Pro.

Using MSL, specialized algorithms that effectively manipulate the depth maps generated by the iPhone 15 Pro's LiDAR camera were implemented. These depth maps are crucial for various image processing applications, including advanced photographic techniques that require a detailed understanding of the scene's geometry.

In this study's setup, the iPhones serve as the primary data collection devices due to their integrated LiDAR scanners. The captured images are processed by MSL shaders in the GPU then output to a framebuffer managed by the GPU for display on the iPhone screen. The iPhone CPU is only used for coordinating between the user interface and GPU, as well as for loading the background image and configuring the GPU render pipeline before it runs. By leveraging MSL, the software can perform complex operations such as filtering, segmenting, and combining depth information with visual data at high speeds, significantly enhancing the system's responsiveness and the overall user experience.

Furthermore, MSL's integration with Xcode and Swift frameworks provides a seamless workflow for developers. The Xcode IDE, as the backbone of Apple's software development, offers a robust environment where LiDAR data and MSL shaders can be meticulously optimized and debugged.
This synergy between hardware capabilities, software frameworks, and the powerful MSL results in a system that can efficiently manage and exploit the benefits of LiDAR technology.\vspace{0.1cm}

\subsubsection{Xcode} Apple's integrated development environment, Xcode, provides a comprehensive array of tools for the incorporation of LiDAR depth data into complex image processing applications. It also provides powerful CPU and GPU profiling tools for debugging and performance benchmarking.\vspace{0.1cm}

\subsubsection{Apple Frameworks} We leveraged several powerful frameworks provided by Apple that are especially adept at managing the video and depth data collected by the iPhone 15 Pro, facilitating high-fidelity image processing. These frameworks include Foundation, AVFoundation, MetalKit, Metal, SwiftUI, CoreImage, CoreMedia, CoreVideo, Photos, Combine, Accelerate, and SIMD.\vspace{0.1cm}

\subsubsection{OpenCV \& Open3D} OpenCV was utilized during prototyping for its advanced image processing capabilities, while Open3D was implemented to manage and visualize 3D data. These libraries play an important role in the interpretation and manipulation of LiDAR-derived depth information, supporting a broad spectrum of applications from augmented reality to environmental modeling. Python's scripting and algorithm development capabilities complement this platform, facilitating the efficient processing and nuanced analysis of LiDAR data.\vspace{0.1cm}

\subsection{Laser Safety Check}
A safety check of the LiDAR sensor of an iPhone 15 Pro Max was performed for background information. The iPhone 15 Pro LiDAR specs are not published by Apple aside from the fact that it contains a Class 1 laser \cite{iphoneuserguide2024b} . Independent parties have acquired more detailed specifications. A blog on the DigiKey website mentioned that the laser operated at 940nm \cite{shepard2022}.

\begin{figure}[!t]
\centering
\includegraphics[width=3.2in]{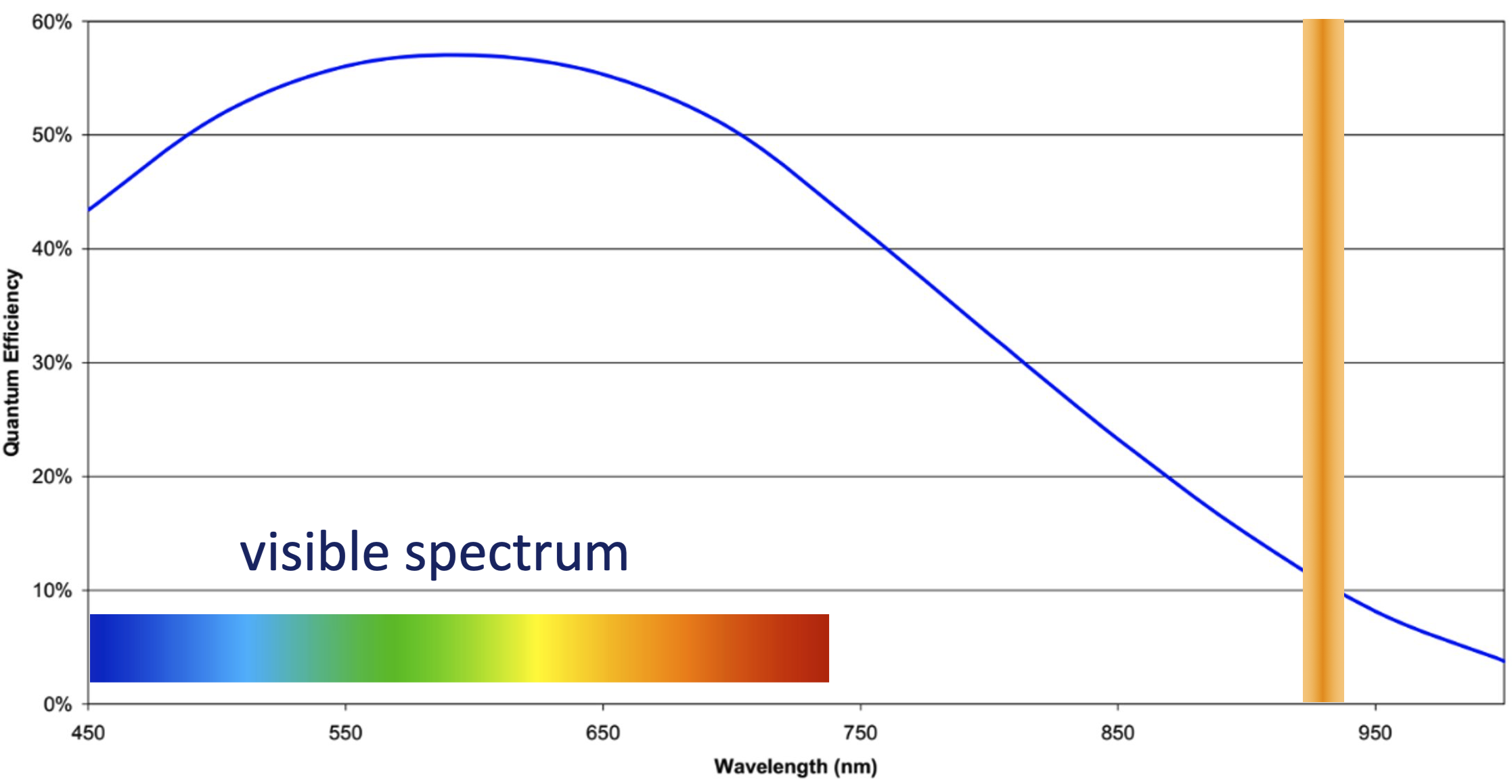}
\caption{The quantum efficiency (QE) of the BMCC \cite{fairchilddatasheet2024} sensor relative to the 940nm wavelength of the iPhone LiDAR laser.}
\label{qe_graph}
\end{figure}

\begin{figure}[!t]
\centering
\includegraphics[width=3.2in]{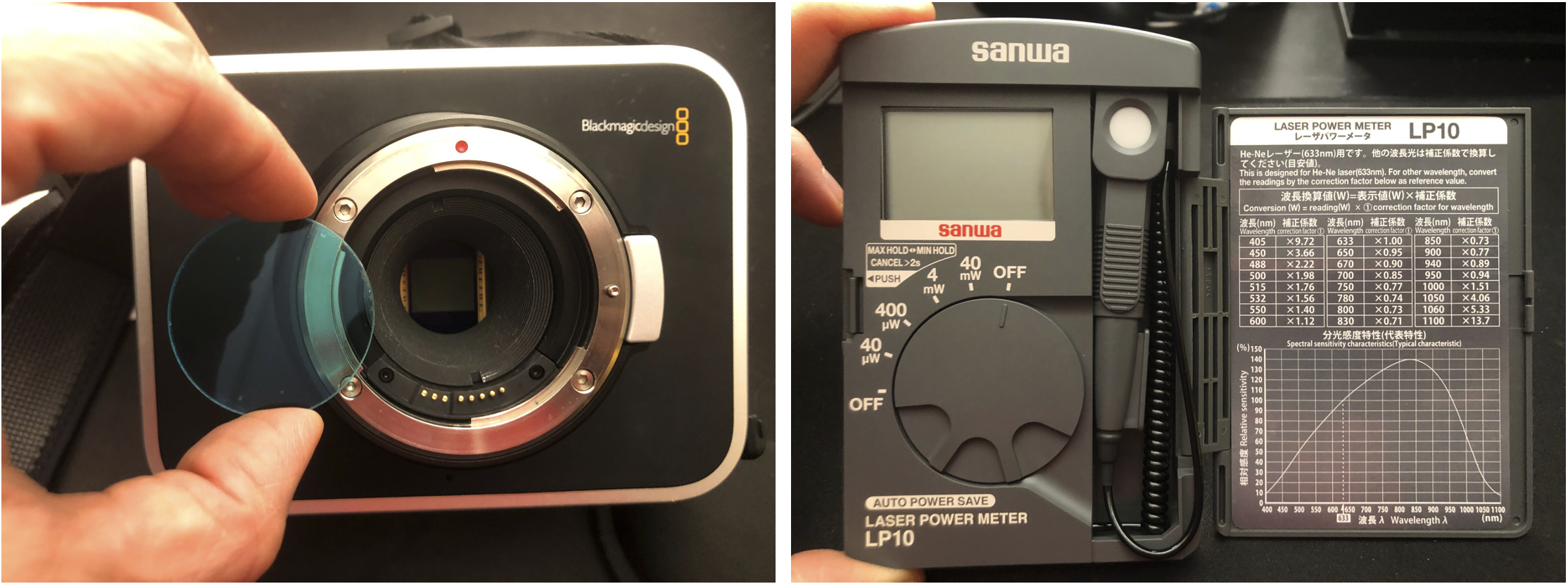}
\caption{BMCC with IR cut filter removed (left). Sanwa LP10 Pocket Power Meter (right).}
\label{bmcc_sanwa}
\end{figure}

\begin{figure}[!t]
\centering
\includegraphics[width=3.2in]{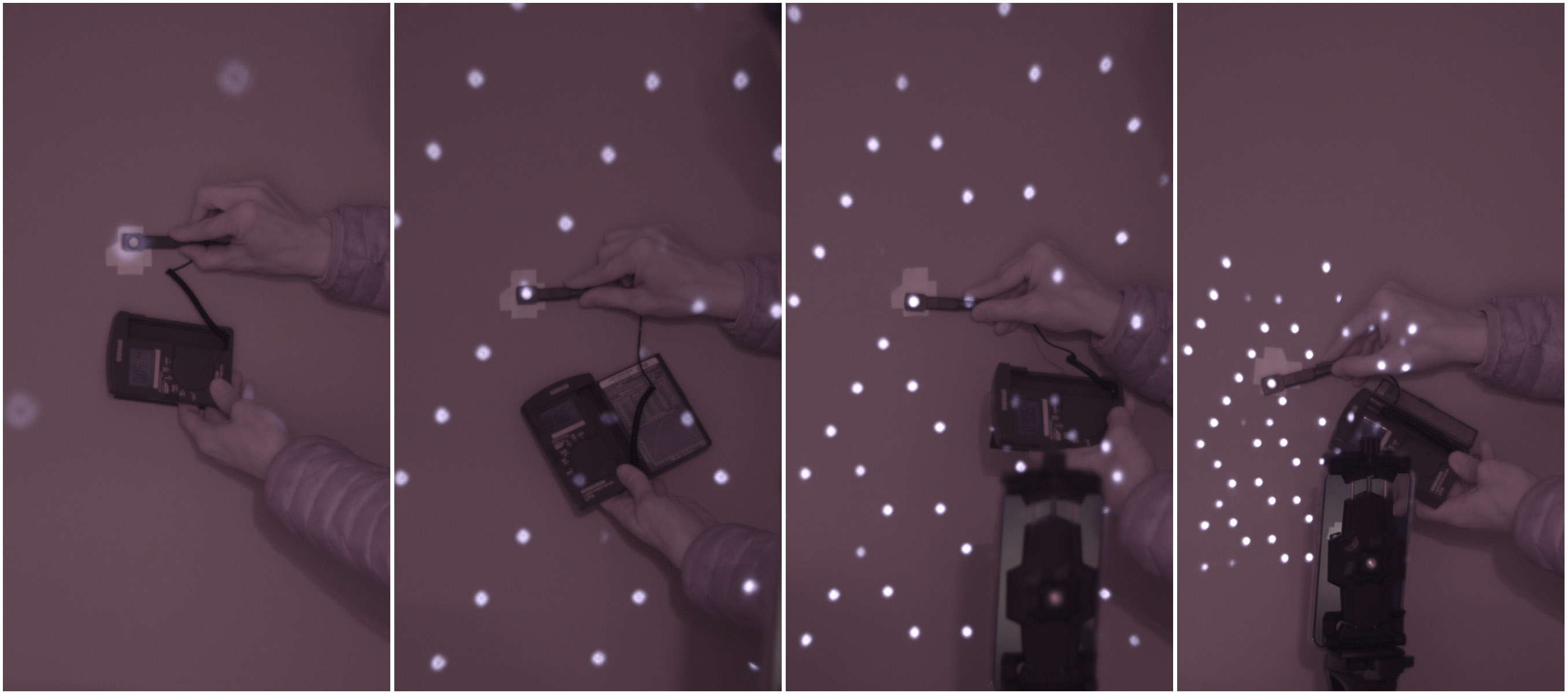}
\caption{Power measurement of the iPhone 15 Pro Max LiDAR laser dots at varying distances.}
\label{laser_dots}
\end{figure}

The Blackmagic Cinema Camera 2.5K (BMCC) was used to visualize the LiDAR laser dots while performing measurements. The BMCC uses a Fairchild Imaging CIS252F sensor \cite{blackmagic2014} that has sensitivity to the 940nm range \cite{fairchilddatasheet2024} as shown in Fig. \ref{qe_graph}, which can be maximized by removing the included IR cut filter as shown in Fig. \ref{bmcc_sanwa}. Using the camera with the IR filter removed enabled visualization of the LiDAR pattern via the camera’s LCD display. A laser dot was measured with the Sanwa LP10 Pocket Power Meter at distances of 78, 30, 20, 9 and 0 inches as shown in Fig. \ref{bmcc_sanwa} and Fig. \ref{laser_dots}.

\begin{figure}[!t]
\centering
\includegraphics[width=3.2in]{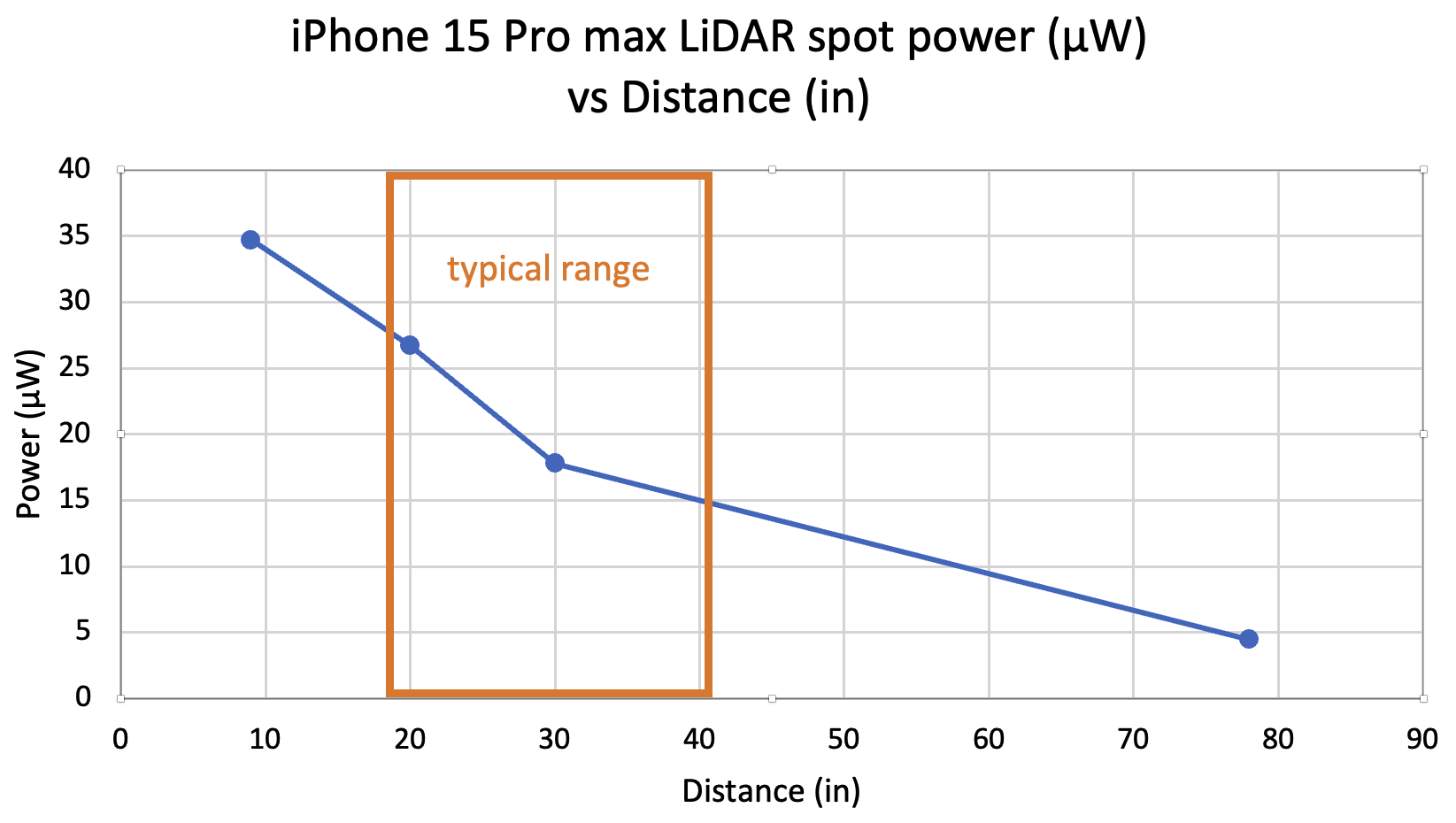}
\caption{The measured power of a given LiDAR dot at varying distances from the power meter. The orange box represents the typical range of a person holding the iPhone at arm length.}
\label{laser_dot_graph}
\end{figure}

In a dimly lit room, the iPhone 15 Pro Max and BMCC were both placed on tripods. The LiDAR laser pattern was projected onto a piece of white tape on a wall so that a laser dot could be centered on the tape for measurement. One person monitored the BMCC display to help the second person line up the laser dot with the power meter. Fig. \ref{laser_dot_graph} shows the laser power measured at each distance, except 0 inches. At 0 inches, the 110 dot matrix was concentrated into a single point on the laser power meter with a maximum power of approximately 4000 µW (4mW).

\begin{figure}[!t]
\centering
\includegraphics[width=3.2in]{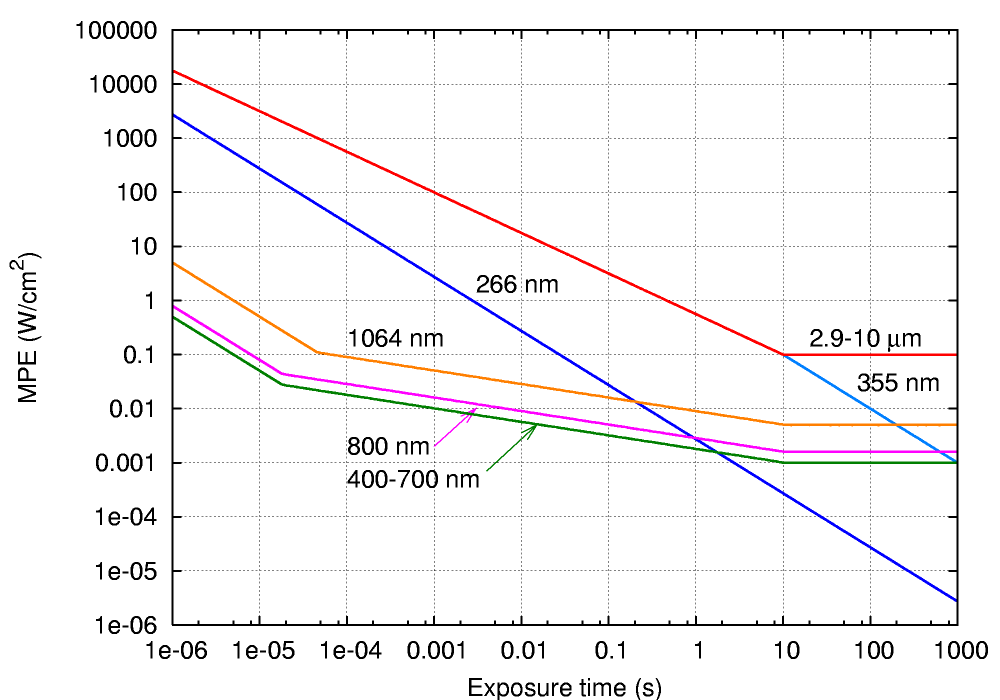}
\caption{The Maximum Permissible Exposure (MPE) chart displays the relationship between power density and exposure duration for different wavelengths, based on the formulas from IEC 60825. This chart was created by Han-Kwang Nienhuys \cite{nienhuys2024}.}
\label{mpe_graph}
\end{figure}

As can be seen in Fig. \ref{mpe_graph}, for a laser in the 940nm range, the maximum permissible (ocular) exposure (MPE) for 10 seconds or more is slightly above 1mW on the logarithmic scale. This implies that the exposure power levels of the iPhone LiDAR sensor, assuming it is used at arm length, is well under the MPE level. It is important to note, however, that the LiDAR should not be used at a distance less than about 5 inches, where more than one dot in the laser projection matrix can enter the eye at once. This distance would put the power exposure to the eye at approximately 4mW.

\section{Methodology}
\subsection{Data set} 
\begin{figure}[!t]
\centering
\includegraphics[width=3.2in]{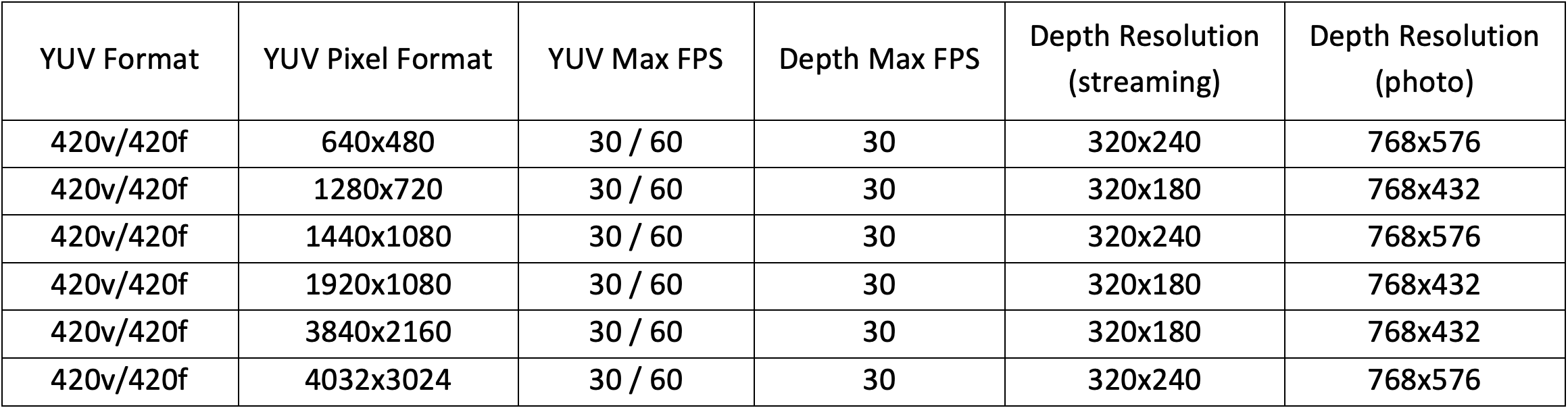}
\caption{iPhone 15 Pro video camera and LiDAR resolutions at varying frame rates \cite{{lidardepthformats2024}}.}
\label{lidar_res_table}
\end{figure}

We created an original dataset for prototyping algorithms by capturing a diverse array of subjects using the advanced capabilities of the iPhone 15 Pro Max. We obtained 20 subject images in varying lighting conditions with subjects of a wide array of skin tones. The iPhone 15 Pro produces LiDAR photos at a higher resolution than LiDAR streamed images. Our prototyping dataset consisted of color photos of 1440x1920 pixels and LiDAR images of 576x768 pixels. By contrast, the final resolution of our iPhone streaming app uses color images of 1440x1920 pixels and LiDAR images of 320x240 resolution streamed at 30 frames per second. Fig. \ref{lidar_res_table} shows the limitations of LiDAR streaming resolution relative to photo resolution. Color images are scaled and cropped by the render pipeline to fit the LiDAR image aspect ratio, and therefore do not match the photo dimensions for maximum color image resolution shown in Fig. \ref{lidar_res_table}. 

\begin{figure}[!t]
\centering
\includegraphics[width=3.2in]{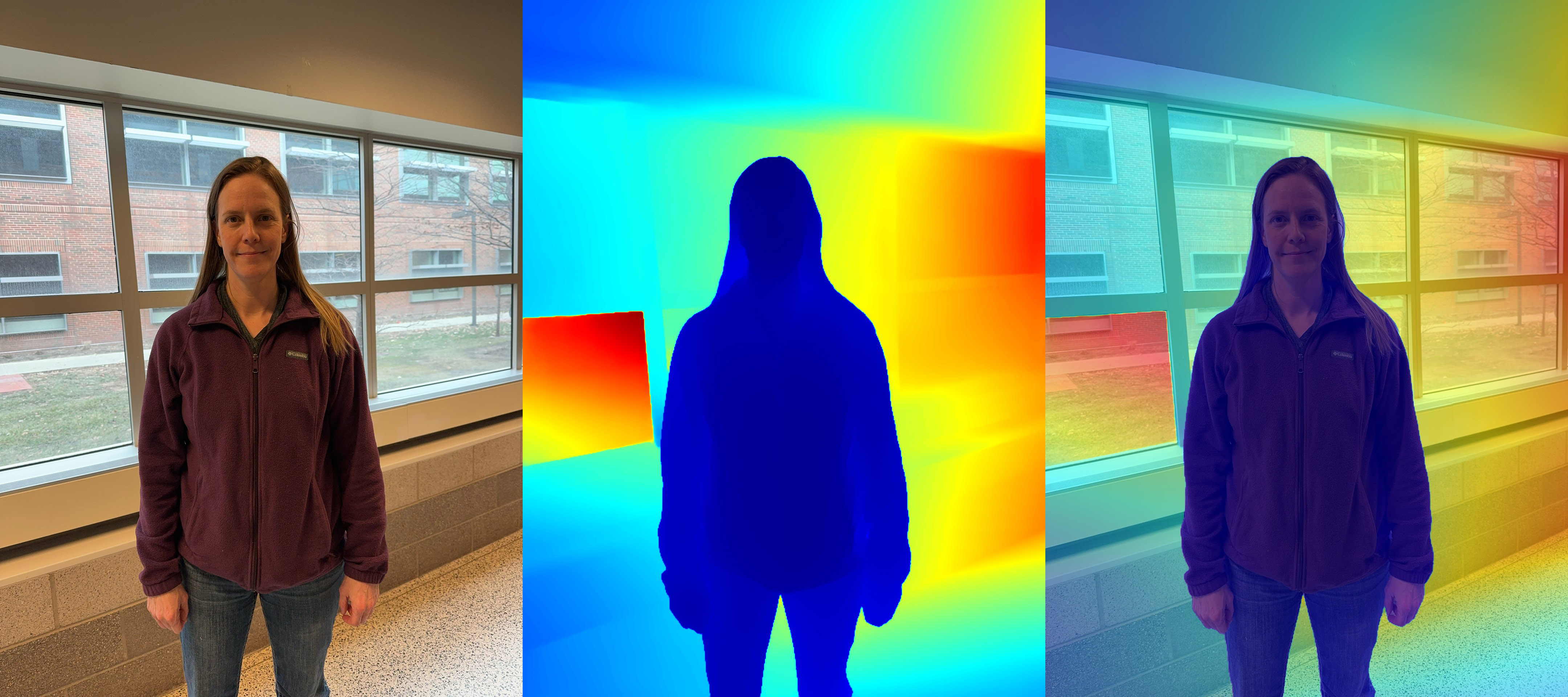}
\caption{Example of prototyping images: (left) 1440x1920 color image, (middle) 576x768 LiDAR image, (right) LiDAR image scaled 2.5x to align with color image.}
\label{prototyping_images}
\end{figure}

The LiDAR images captured as photos for prototyping were scaled by 2.5x via linear interpolation to align with the color images for depth thresholding. Fig. \ref{prototyping_images} shows an example of a prototyping color photo, its associated scaled LiDAR image, and how they align.

\subsection{Data Collection}
An Apple Inc. provided example app Xcode project named “Capturing depth using the LiDAR camera” \cite{appledeveloper2024} was modified by the team to export images in JPEG format synchronized with their corresponding LiDAR data in a binary file containing a list of depth values in 32-bit float format. See Appendix for the specific code modifications.

\begin{figure}[!t]
\centering
\includegraphics[width=3.2in]{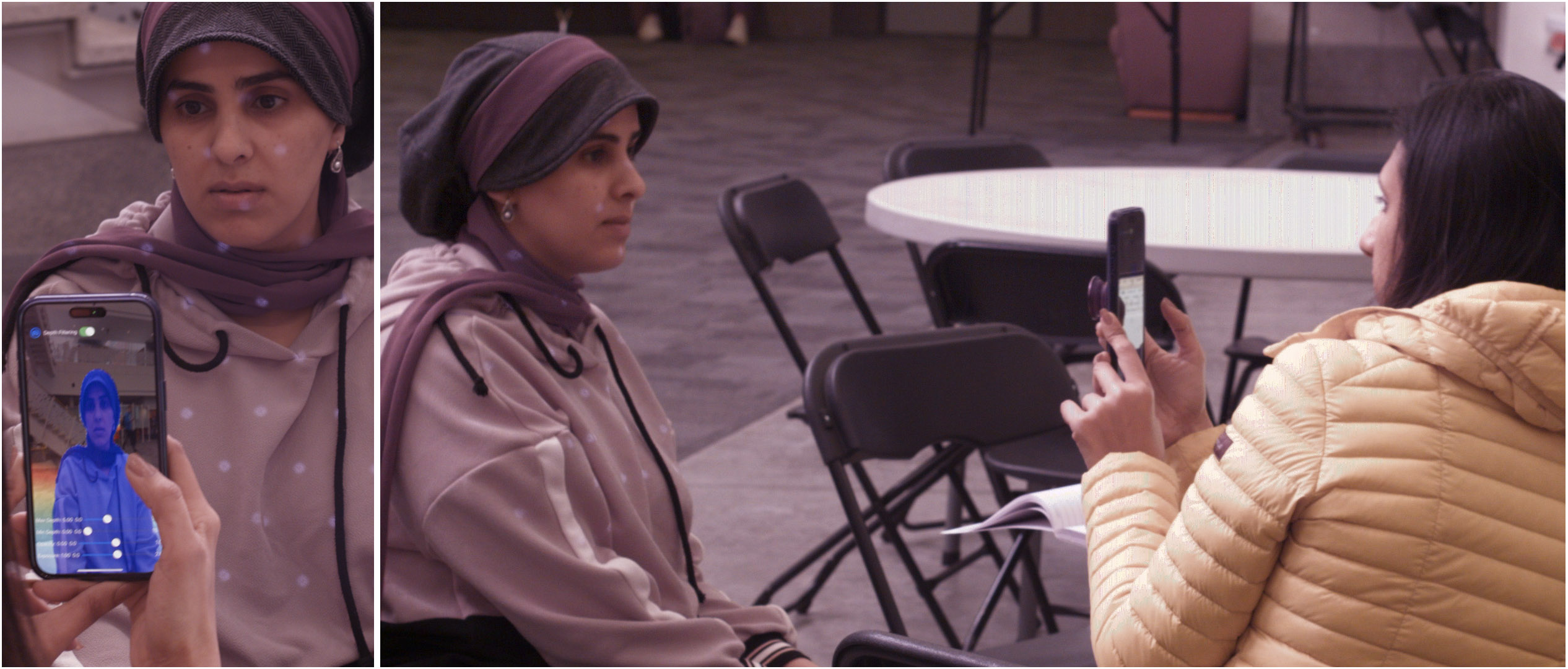}
\caption{Image and LiDAR data captured as observed by the BMCC with IR cut filter removed.}
\label{cut_filter_removed}
\end{figure}

Images and depth data for each subject were captured as shown in Fig. \ref{cut_filter_removed}. A variety of low-light conditions were used for data capture. The URCERI MT-912 light meter was used to measure the illumination level of the scene for each sample image taken. Illumination was measured in lux. The range lighting conditions was from 0 to 75 lux.

The images and binary depth data were then read by a Python script written by the team where they were processed with OpenCV. 

\subsection{Data Flow}
\begin{figure}[!t]
\centering
\includegraphics[width=3.2in]{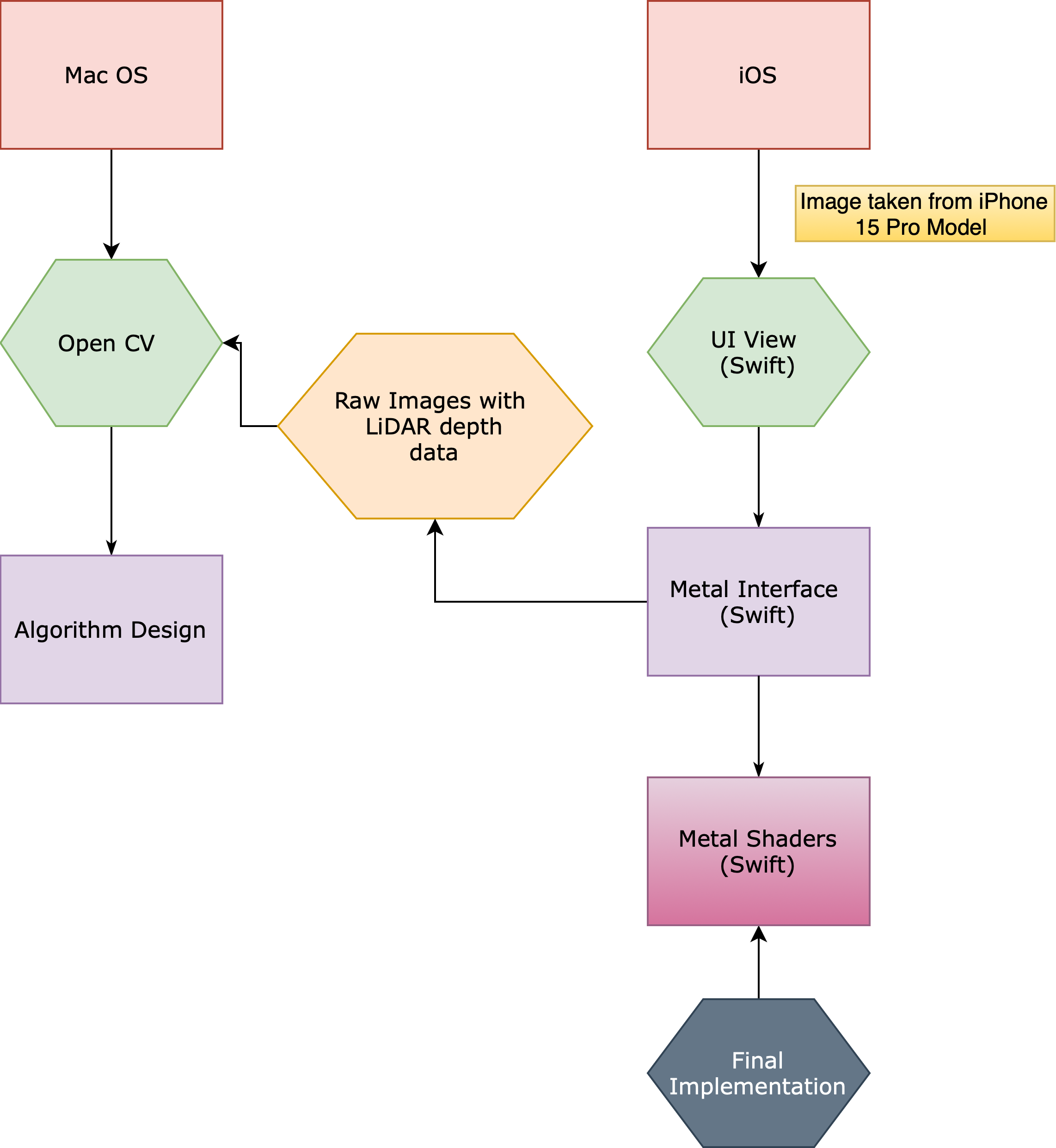}
\caption{Flowchart of the image processing with LiDAR on mobile devices.}
\label{flow_chart_high_level}
\end{figure}

Fig. \ref{flow_chart_high_level} shows a flow chart of the image processing for prototyping on MacOS and final implementation on iOS. Still images and their corresponding LiDAR data were exported from the modified iOS example app and used to prototype the proposed algorithms with Python OpenCV libraries in MacOS. The final algorithms were implemented in Metal Shader Language (MSL) in the shaders.metal file within the modified example app.

\subsection{Algorithms and Data Structures}
Image processing algorithms were prototyped using OpenCV in Python, evaluating their performance in terms of both effectiveness and efficiency. Descriptions of the algorithms prototyped are included below.

\subsubsection{Background Removal} 
\begin{figure}[!t]
\centering
\includegraphics[width=3.2in]{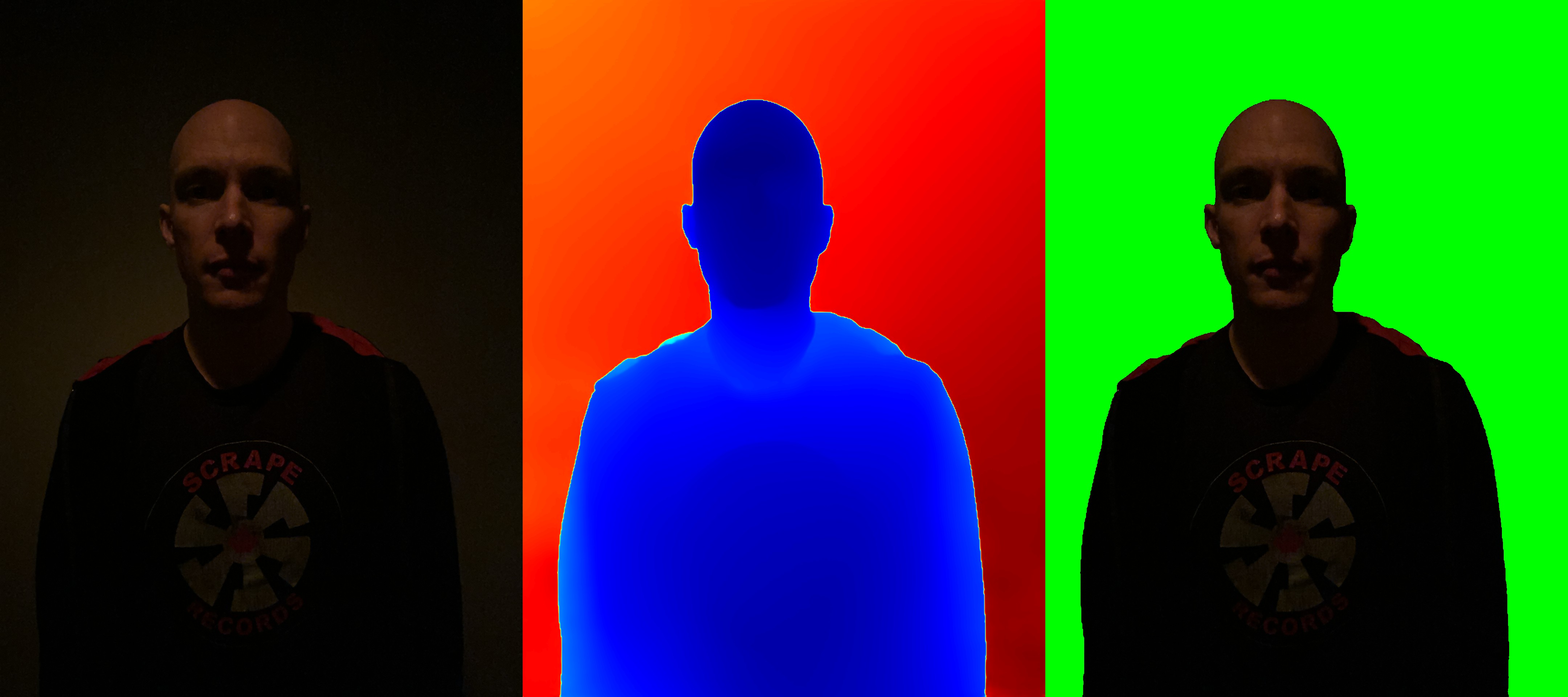}
\caption{Low-light background removal in Python: Original image (left), LiDAR depth map of subject (middle), depth-based background removal (right).}
\label{low_light_green}
\end{figure}

This algorithm uses depth information, which could be obtained from dual-camera systems, depth sensors, or estimated using machine learning models, to distinguish between the foreground subject and the background. By understanding the LiDAR depth and its alignment to a corresponding color image, the algorithm can accurately remove the background while preserving the subject (Fig. \ref{low_light_green}). In prototyping, the iPhone color image we obtained from the iPhone 15 Pro Max was converted to RGBA format.  The alpha value of each pixel was set to 0 or 1 if the depth value associated with the pixel in the companion depth map was above or below the depth threshold, respectively.
 
\subsubsection{Exposure Boosting} 
Exposure boosting using gamma correction is a technique used in digital image processing to enhance the visibility of details in pictures, especially in areas that are underexposed or have poor lighting conditions. Gamma correction manipulates the luminance of the pixels in an image in a non-linear way, which can be particularly effective in correcting exposure without compromising the overall balance of light and dark areas in the image.
Gamma correction adjusts the brightness of an image by changing its pixel values using a special rule based on powers. This makes some parts of the image lighter or darker to improve overall visibility. 
If $\gamma<1$: This brightens the image, which is useful for dark images where details are lost in shadows.
If $\gamma>1$: This darkens the image, helpful for overexposed images where highlights need to be controlled.

\subsubsection{Noise Reduction}
\begin{figure}[!t]
\centering
\includegraphics[width=3.2in]{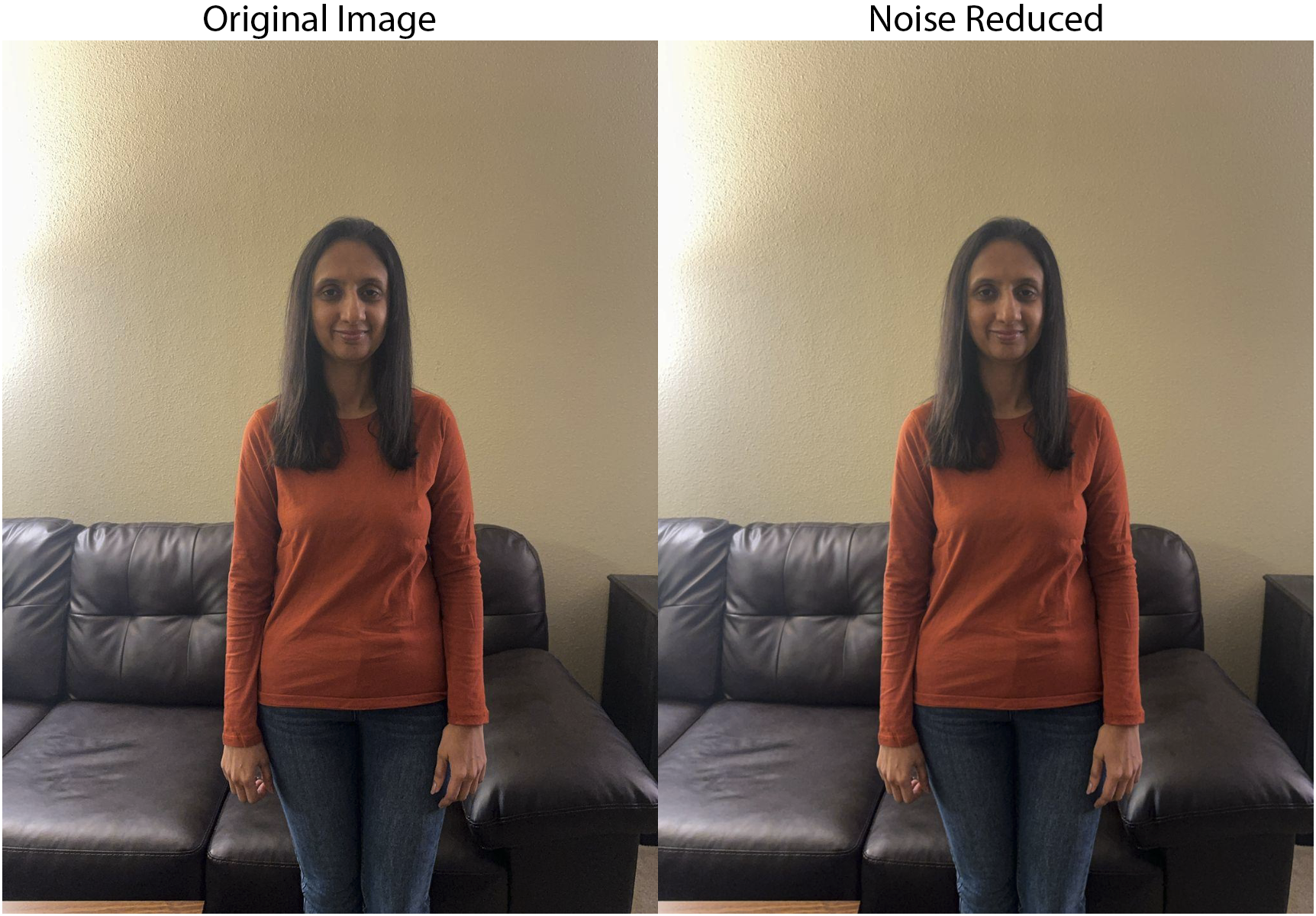}
\caption{Result of prototype noise reduction algorithm in Python.}
\label{prototype_noise_reduction}
\end{figure}

During the prototyping phase, multiple noise reduction techniques available in OpenCV were explored to determine their efficacy in reducing noise artifacts while maintaining image integrity. Methods such as Gaussian blur, median blur, and bilateral filtering were explored. Gaussian blur is commonly used for its simplicity and effectiveness in reducing image noise and smoothing details, but it can blur edges, which might not be desirable in all scenarios. Median blur is effective in removing salt-and-pepper noise and is better at preserving edges compared to Gaussian blur. Bilateral filtering is more sophisticated, as it reduces noise while preserving edges by using a non-linear filter that considers both the difference in pixel intensities (range filter) and the geometric closeness of pixels (spatial filter).
The integration of these methods into a single, coherent noise reduction strategy involved iterative testing to balance noise removal with detail preservation (Fig. \ref{prototype_noise_reduction}). The algorithm's performance was evaluated on a series of test images captured under various lighting conditions to simulate real-world use cases. Adjustments were made based on the feedback from these tests, focusing on optimizing parameters such as the kernel size for the blurs and the diameter, sigmaColor, and sigmaSpace for the bilateral filter.
Furthermore, OpenCV’s support for Python made the prototyping phase highly efficient due to Python’s simplicity and the ability to quickly write and execute code. This allowed for fast iterations and real-time testing, which is crucial for fine-tuning the algorithm to achieve the desired balance between noise reduction and image quality. 

\subsubsection{Color Rebalancing} 
\begin{figure}[!t]
\centering
\includegraphics[width=3.2in]{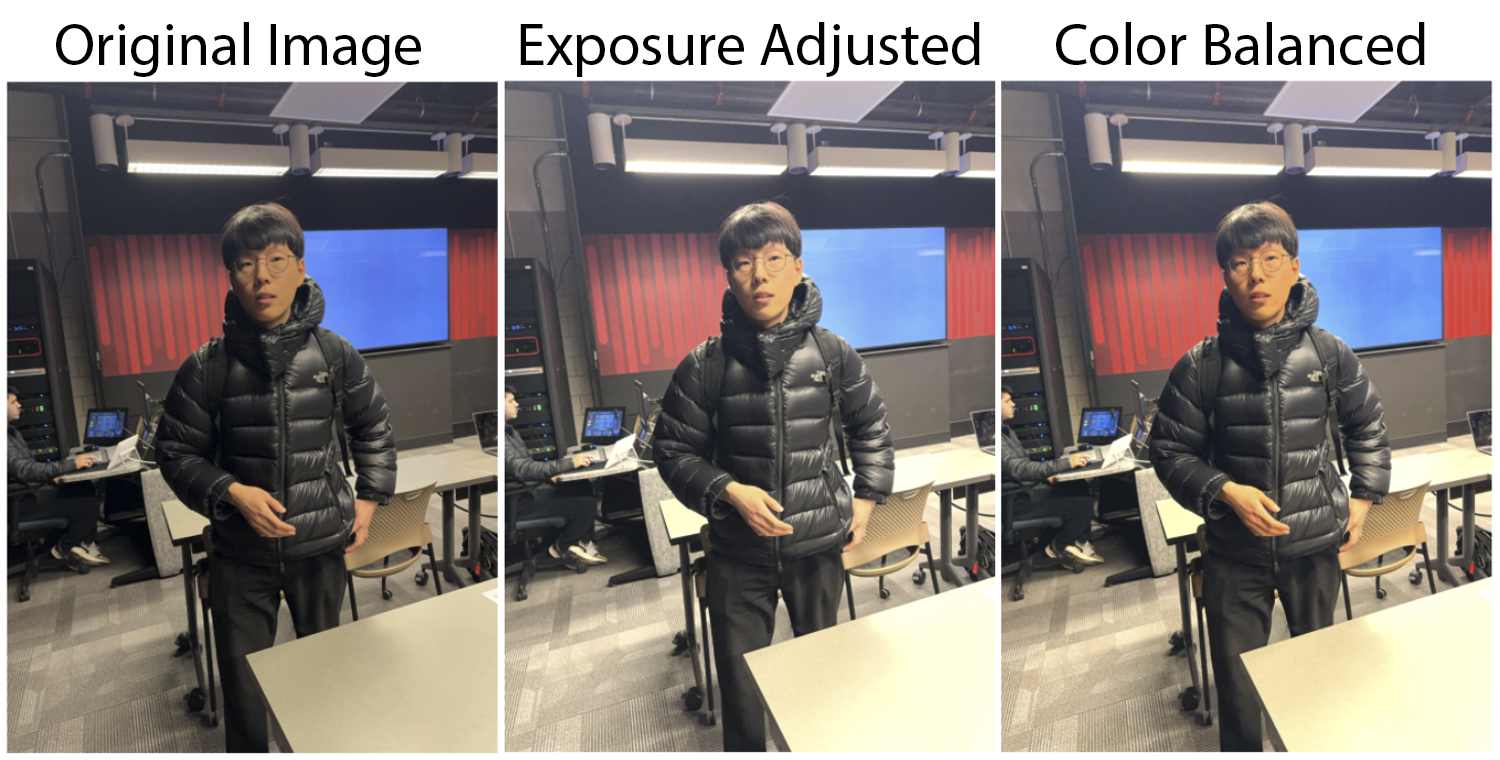}
\caption{Result of prototype color balancing algorithm.}
\label{prototype_color_balance}
\end{figure}

For implementing the color rebalancing prototype, images obtained from the LiDAR depth data was carried out using Python, leveraging the powerful OpenCV library for image manipulation tasks and the Matplotlib library for visualization purposes. The prototype is designed to enhance images through two primary techniques: exposure adjustment and color balance optimization. The exposure adjustment function utilizes parameters alpha and beta to control the gain and brightness of the image respectively. Following exposure adjustment, the prototype uses a color balancing method, which adjusts the saturation of the image. This is particularly beneficial for images that appear washed out or under-saturated, breathing life back into the visual content (Fig. \ref{prototype_color_balance}).

\subsubsection{Edge Smoothing}
\begin{figure}[!t]
\centering
\includegraphics[width=3 in]{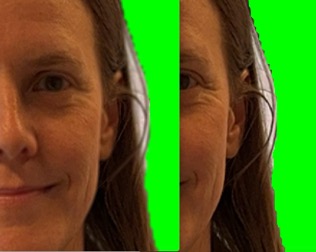}
\caption{Result of prototype edge smoothing algorithm in Python (left) compared to original image (right).}
\label{prototype_edge_smoothing}
\end{figure}

Edge smoothing is a critical step in image processing, particularly when preparing images for higher-level tasks like object detection or segmentation. It helps in reducing noise and improving the visual quality of edges in digital images. A combination of morphological operations—specifically erosion and dilation—followed by Gaussian blur is used to achieve effective edge smoothing.
The prototyping for this feature was carried out by a combination of edge detection and morphological operations to enhance the edges within an image, utilizing the capabilities of the OpenCV library. Specifically, the script reads image in BGRA format and converts it to BGR to discard the alpha channel. The main functionality begins with the detection of edges using the Sobel operator applied in both x and y directions to the grayscale conversion of the image. The resulting edge magnitudes are calculated to produce a comprehensive edge map of the original image. This edge map is then thresholded to create a binary edge image where the edges are distinctly separated from non-edge areas.
Following the edge detection, the script performs morphological operations, specifically erosion and dilation on the thresholded edge map. Erosion helps in reducing the noise and the size of the edges, which helps in removing finer, unwanted details, while dilation follows to expand the refined edges, potentially reconnecting fragmented parts to enhance continuity and visibility. These processed edges are then combined back onto the original image to accentuate the edges clearly, using a mask created from the dilated edges. Finally, a comparison plot is generated that visually contrasts the original and edge-enhanced images (Fig. \ref{prototype_edge_smoothing}). This visualization is crucial for evaluating the effect of the edge enhancement, offering a clear before-and-after perspective on the applied image processing techniques.

\subsubsection{Implementation of the algorithm in MSL} 
\begin{figure}[!t]
\centering
\includegraphics[width=3.2in]{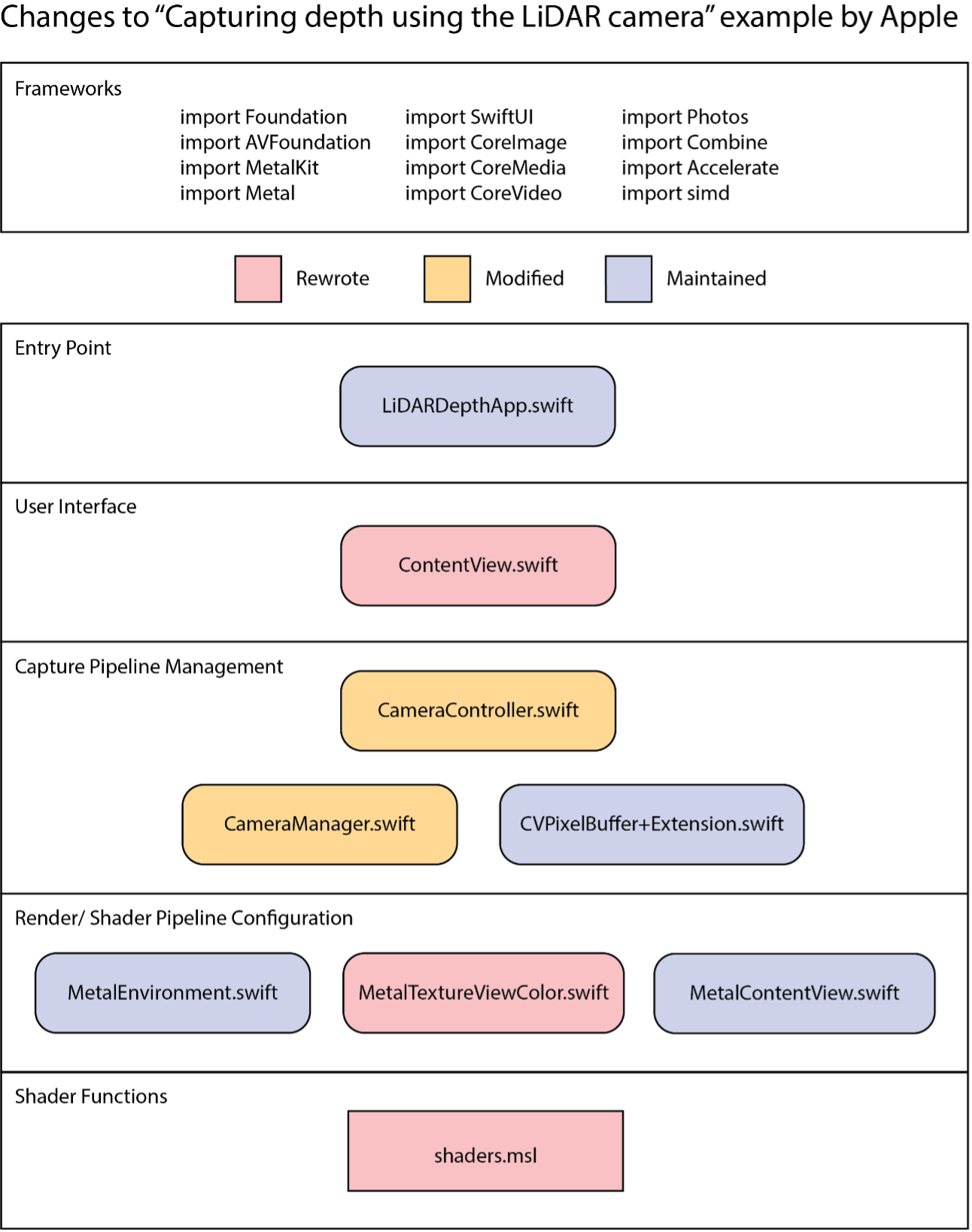}
\caption{Components modified in the app “Capturing Depth using the LiDAR-by Apple.”}
\label{modified_components_flow_chart}
\end{figure}

\begin{figure}[!t]
\centering
\includegraphics[width=3.2in]{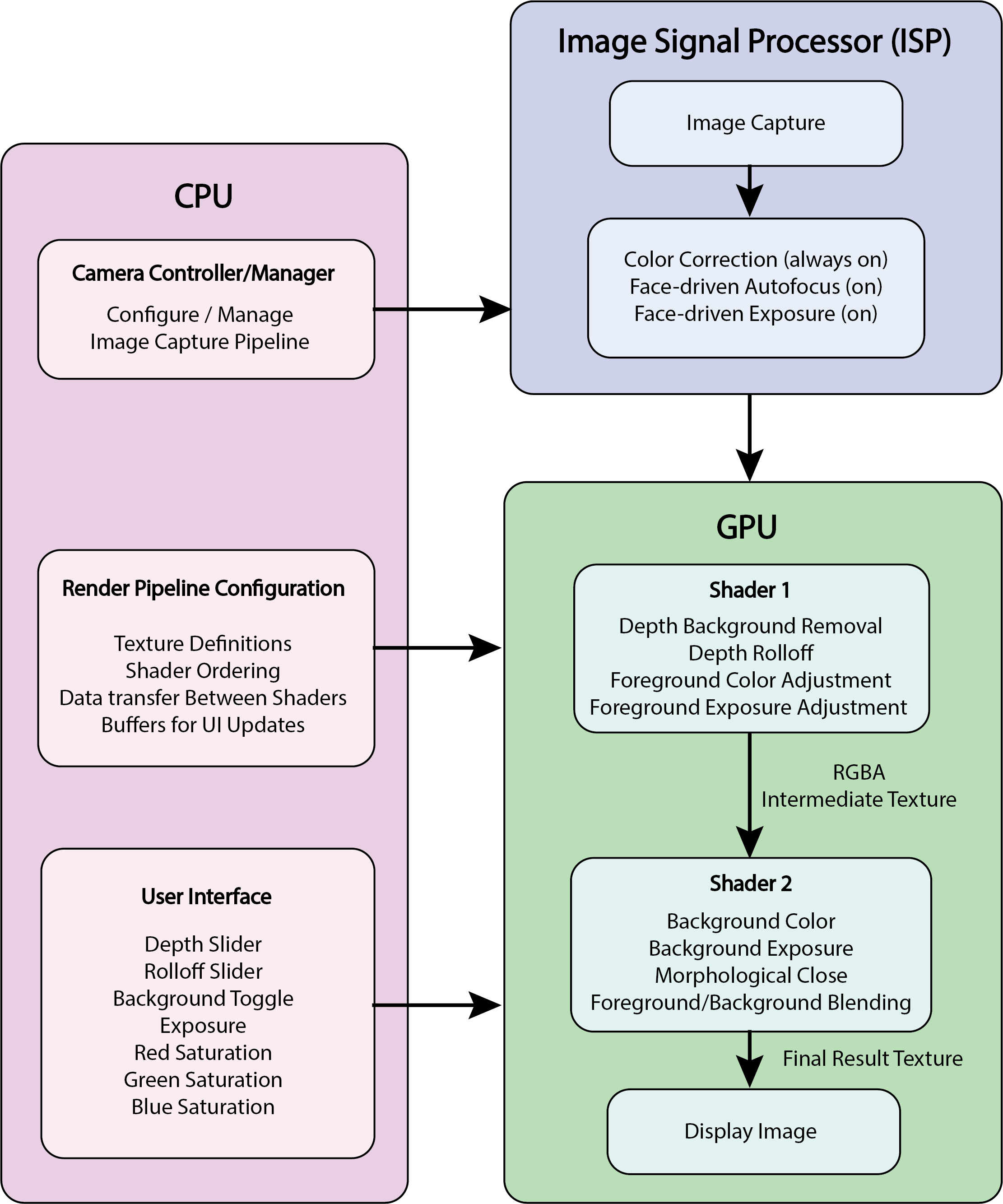}
\caption{Data flow in study's iPhone app.}
\label{data_flow_chart}
\end{figure}

Fig. \ref{modified_components_flow_chart} shows the components of the Apple example project that were rewritten, modified, or left unchanged as part of this study. The original single-shader \textit{shaders.msl} file is almost entirely rewritten to implement the two shaders described in Fig. \ref{data_flow_chart}. Shader 1 includes background removal, color adjustment to the remaining foreground pixels, and setting the alpha channel values for the depth roll-off. The resulting image data is then processed in Shader 2 to compute the morphological close (edge smoothing) operation, and to adjust exposure, color, and blending of the background. These processing steps cannot be performed in a single shader due to the morphological close, which calculates the data in a pixel based on data in surrounding pixels. Since each step in the shader occurs on a single pixel before moving on to the next pixel, a race condition occurs with the morphological close when a neighboring pixel has not yet been computed.

\section{Results}

\subsection{Implementation of Prototyped Algorithms}

\subsubsection{Exposure Boosting}
\begin{figure}[!t]
\centering
\includegraphics[width=3.2in]{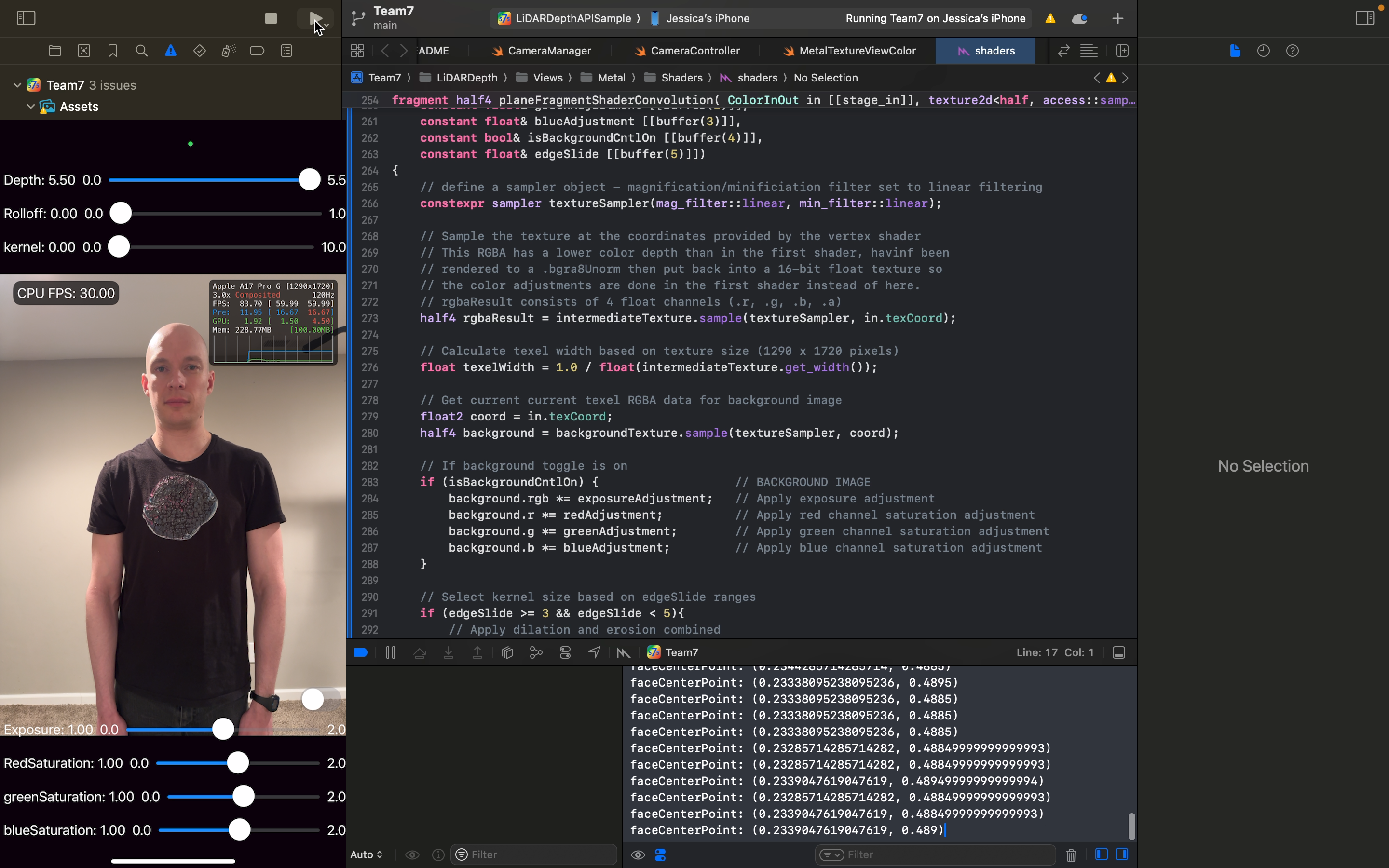}
\caption{Face center point detection printed to the console with face-driven autofocus and auto exposure enabled.}
\label{face_detection}
\end{figure}

\begin{figure}[!t]
\centering
\includegraphics[width=3.2in]{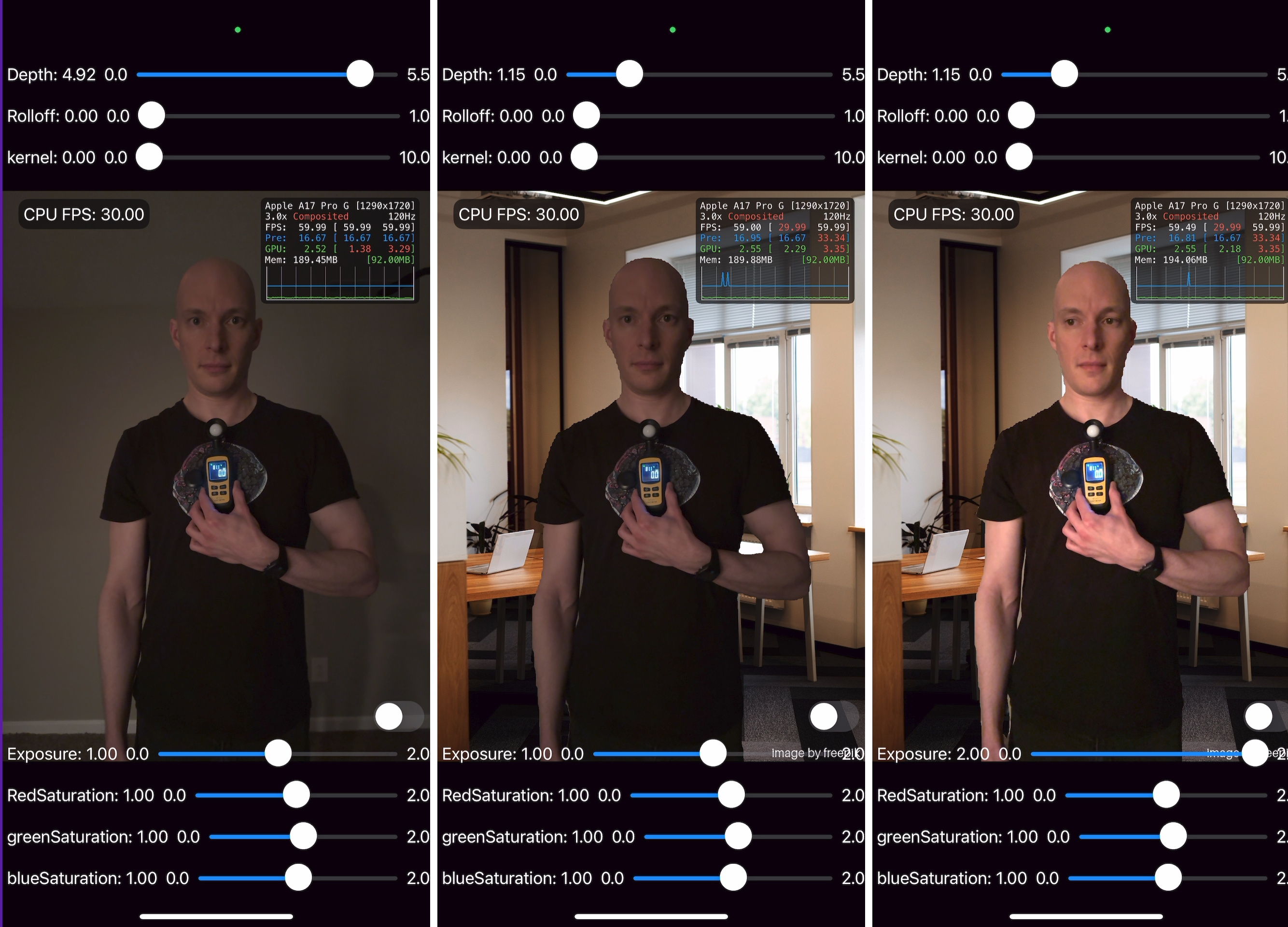}
\caption{Results with various shader algorithms enabled: (left) unprocessed video capture, (middle) video with depth thresholded background removal, (right) video capture with background removal and exposure boost. All images were captured in a $<3$ lux low light setting.}
\label{shader_algorithms}
\end{figure}

The image signal processor (ISP) in the iPhone 15 Pro Max has the ability to perform autofocusing and auto exposure for a general image, or based on a person’s face \cite{avfoundation2024}. We chose to employ the face-driven features in our app by enabling the associated property (Fig. \ref{face_detection}). The auto exposure, however, was not sufficient for more extreme low light environments, so we added a slider to the app to allow the user to boost the exposure by up to 3x auto exposure level (Fig. \ref{shader_algorithms}.

\begin{figure}[!t]
\centering
\includegraphics[width=3.2in]{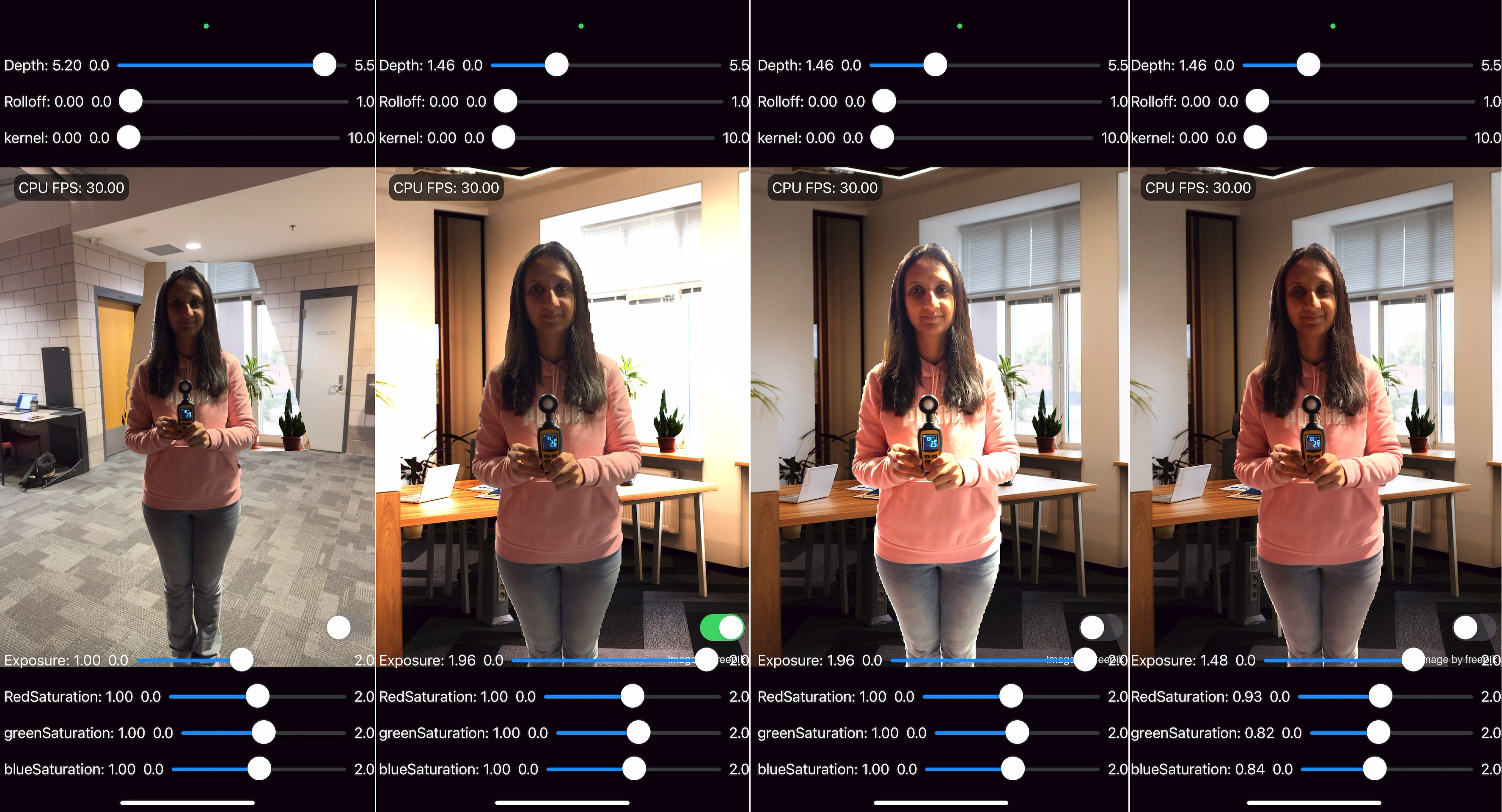}
\caption{Image before depth thresholding (left), over-exposed background (left-center), over-exposed foreground (right-center), properly exposed foreground and background with slight color correction (right).}
\label{exposure_color}
\end{figure}

\subsubsection{Noise Reduction}
As with exposure boosting, noise reduction done by the ISP performed well even with additional exposure boosting.

\subsubsection{Color Rebalancing}
The ISP performed well at automatically correcting color after boosting, however, we implemented sliders on the user interface for manually adjusting the color gain of each color channel in the final image. These sliders are particularly useful for matching the foreground subject to the background image (Fig. \ref{exposure_color}). A toggle was added to the user interface to allow for switching the exposure and color adjustments between the foreground and background image data.

\subsubsection{Edge Smoothing}
Edge smoothing was performed with a morphological close operation on the alpha channel of the foreground image. This consists of a dilation followed by an erosion using the same kernel size. We implemented kernels sizes of 3x3, 5x5, 7x7, and 9x9 as well as the ability to disable smoothing.  A continuous slider on the user interface was used for simplicity. iOS sliders do not inherently support discrete integer values, so any value in [0,3) disables the smoothing, a value in [3,5) enables the 3x3 kernel, and so on. 

\begin{figure}[!t]
\centering
\includegraphics[width=3.2in]{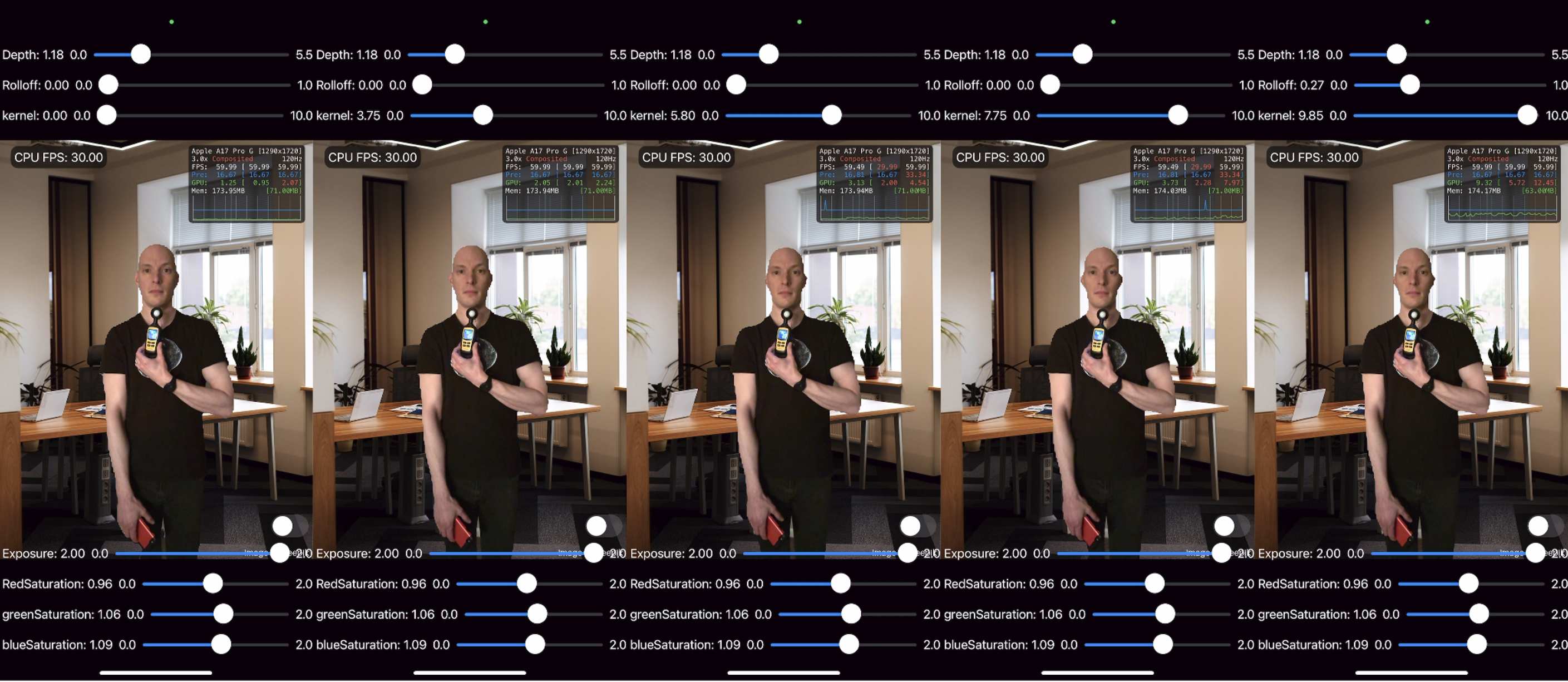}
\caption{Morphological close on alpha of foreground image with various kernel sizes. From left to right: none, 3x3, 5x5, 7x7, and 9x9. See Fig \ref{morph_closeup} for a close-up view.}
\label{morph_close}
\end{figure}

\begin{figure*}[!t]
\centering
\includegraphics[width=6in]{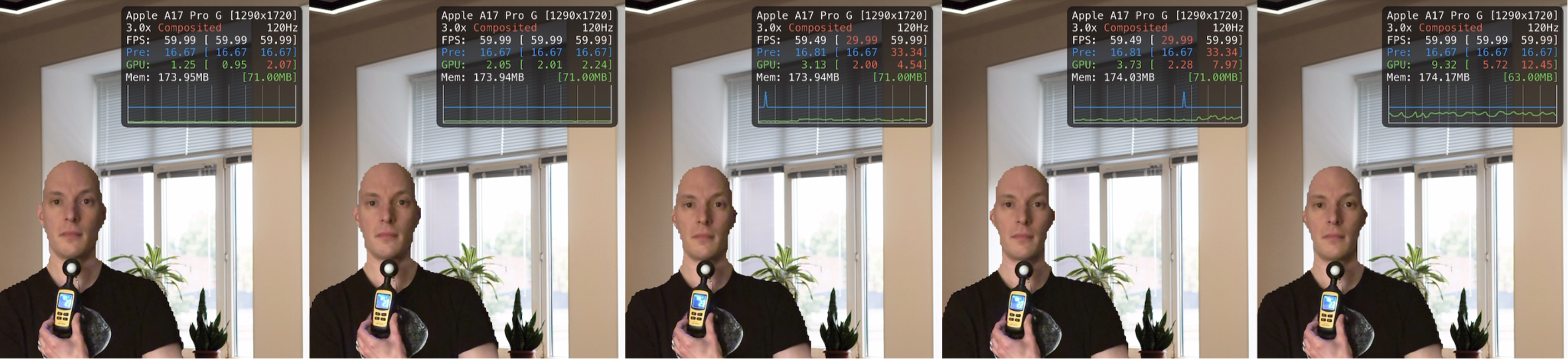}
\caption{Close-up of images from Fig. \ref{morph_close}. Note the detail in the ears being lost as kernel size increases.}
\label{morph_closeup}
\end{figure*}

Implementing this feature was particularly challenging because it required using a second sequential shader and an intermediate texture with the background already removed. This feature could not be implemented in Shader 1 where the background was in the process of being removed because each shader can only operate on individual pixels of an existing texture. The solution was to render Shader 1 to an intermediate texture in an RGBA format which contained the foreground data with alpha channel already adjusted for background removal, then perform the convolution steps of the morphological close operation on the intermediate texture in Shader 2, followed by a mixing operation to combine the alpha-smoothed texture with the background image. This required setting up a custom render pipeline to configure the sequential shaders with the intermediate texture (Fig. \ref{morph_close}, \ref{morph_closeup}).

\subsubsection{Alpha Roll-off}
\begin{figure}[!t]
\centering
\includegraphics[width=3.2in]{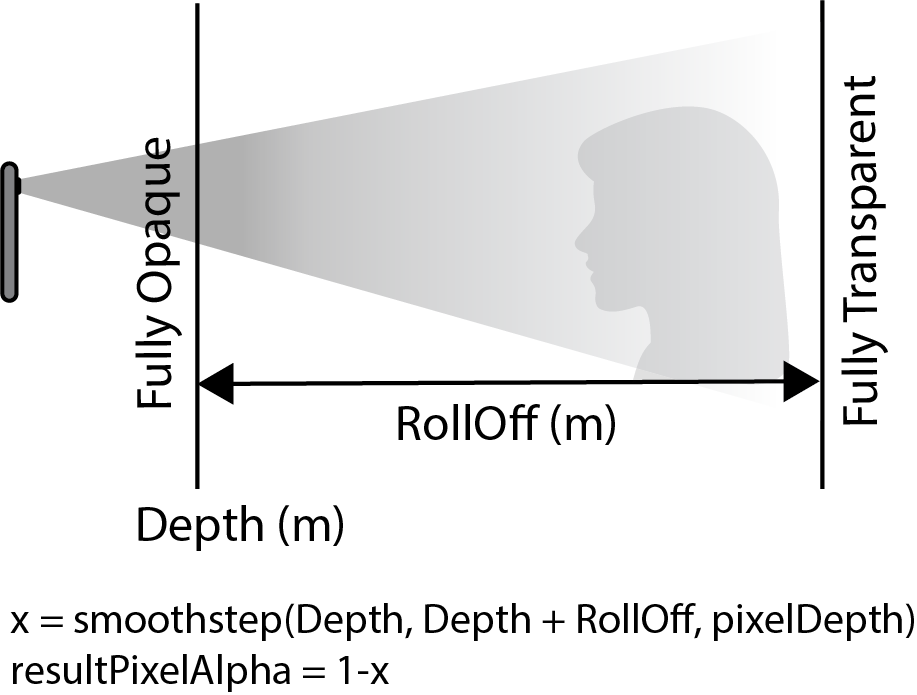}
\caption{Diagram showing how the alpha value of each pixel is calculated when using roll-off.}
\label{roll_off_diagram}
\end{figure}

\begin{figure}[!t]
\centering
\includegraphics[width=3.2in]{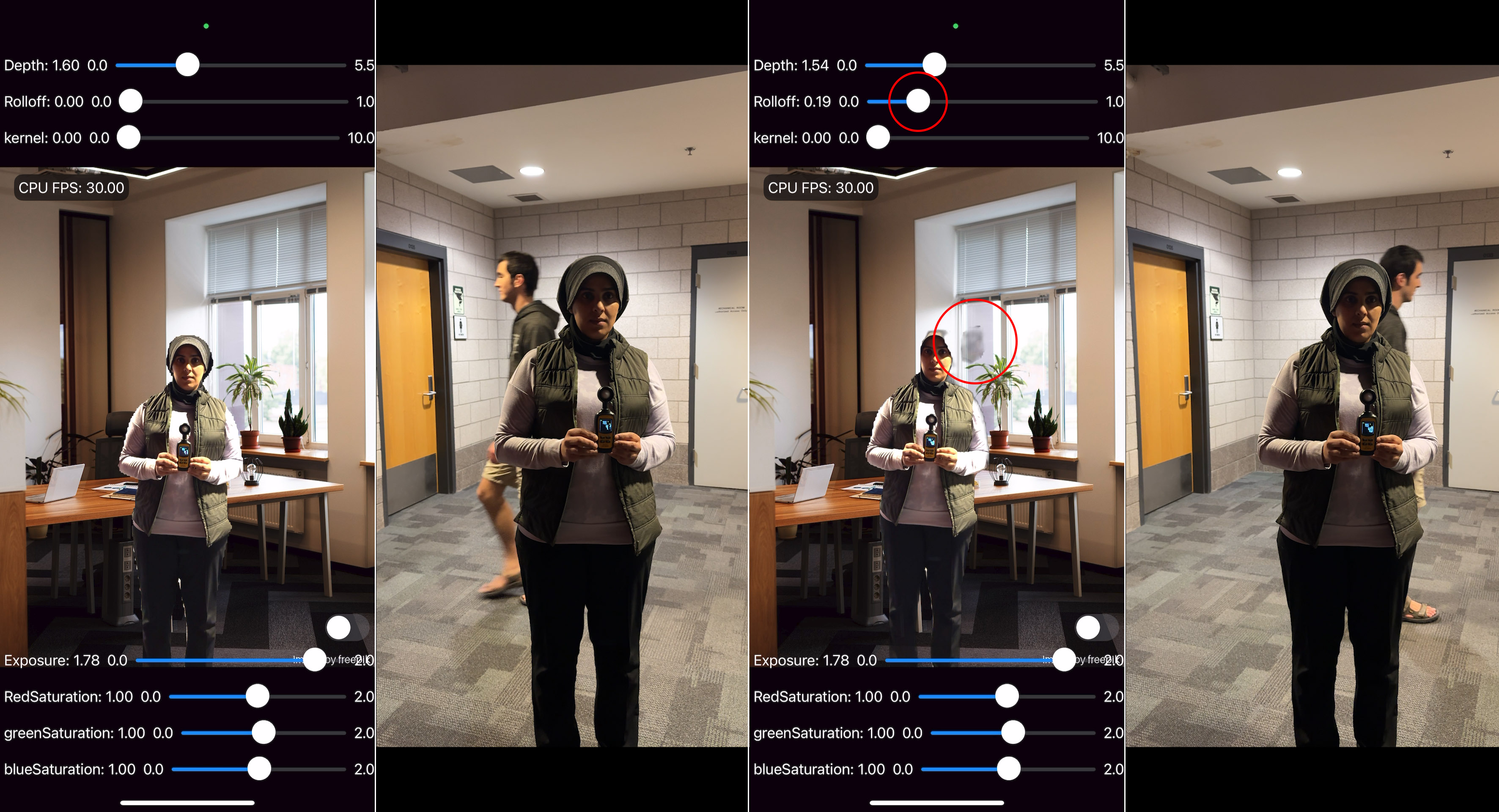}
\caption{Foreground subject isolated from subject in background when Rolloff = 0 (left), slight bleed of background subject into frame when RollOff = 0.19. (right)}
\label{roll_off_bleed}
\end{figure}

\begin{figure}[!t]
\centering
\includegraphics[width=1.8in]{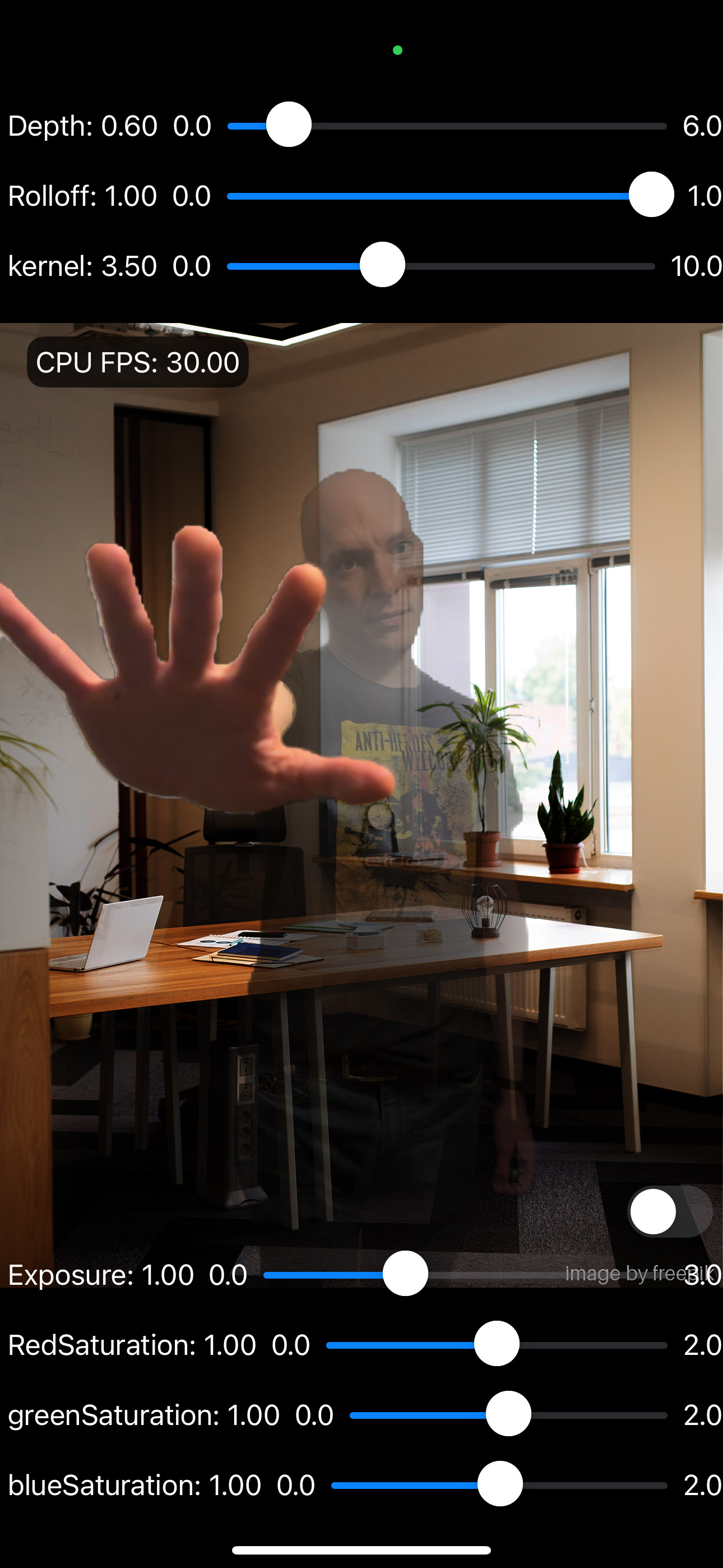}
\caption{Alpha edge roll-off maximized to 1.0 showing depth thresholding as a gradient.}
\label{roll_off_single}
\end{figure}

We added the ability to choose more than just a 1 or 0 for the alpha channel by implementing an optional smooth transition between two depths boundaries. A slider was added to the user interface labeled ``RollOff" that allows the user to select a float value in [0, 1] in meters (implemented in Shader 1). This value determines the size of of the depth transition range (Fig. \ref{roll_off_diagram}). The alpha value of each pixel is calculated using the ``Depth" and ``RollOff" sliders on the user interface. The smoothstep function in MSL performs a cubic interpolation from the fully opaque threshold (Depth) to the fully transparent threshold (Depth + RollOff). This algorithm produces a softening effect that can supplement edge smoothing when used sparingly and produces a stylized effect on the foreground subject when used more liberally (Fig. \ref{roll_off_bleed}, \ref{roll_off_single}).

\subsubsection{Performance Metrics}
The implemented algorithms operate within the 60 frames per second (fps) screen refresh rate. This performance metric is crucial to ensure a seamless user experience during live video streaming.
 
\begin{figure}[!t]
\centering
\includegraphics[width=3.2in]{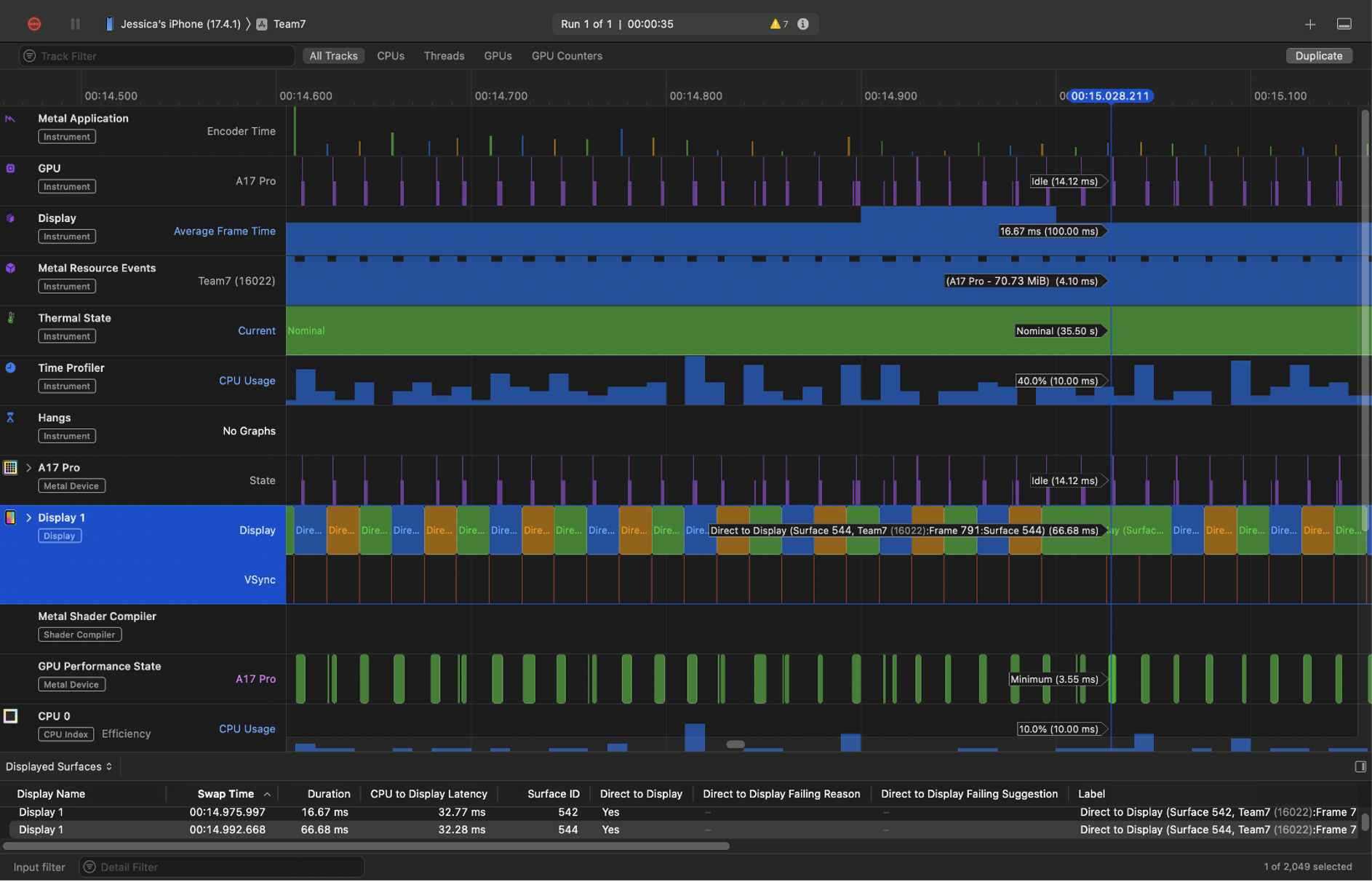}
\caption{Instruments profiler used in Xcode for profiling GPU and CPU usage of the app.}
\label{profiler}
\end{figure}  

Using the Instruments profiler in Xcode (Fig. \ref{profiler}), we determined the execution times of both shaders and found that Shader 1 consistently executed in less than 1ms. The provided data depicted in Table 1 shows the execution times for Shader 2 on a GPU with varying kernel sizes. As expected, larger kernels result in longer processing times. Specifically, bypassing the morphological close operation results in Shader 2 executing in 897.88 microseconds. As the kernel size increases from 3x3 to 9x9, the execution times rise from 1.54 milliseconds to 6.55 milliseconds. The choice of kernel size therefore represents a trade-off between computational speed and the effectiveness of the processing. Conversely, for applications where output quality is more critical, larger kernels might be warranted despite the higher computational cost.

\begin{figure}[!t]
\centering
\includegraphics[width=3.2in]{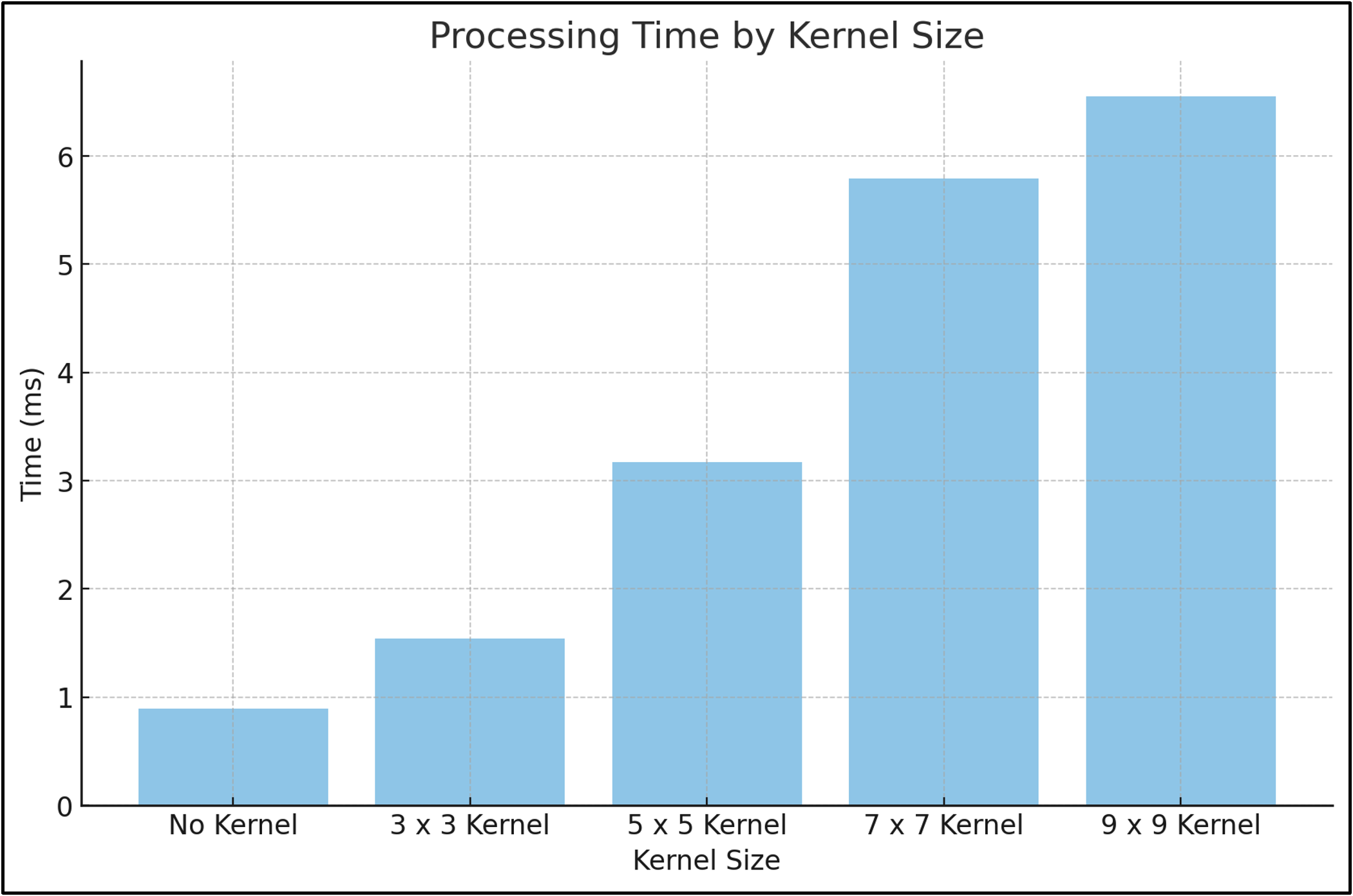}
\caption{Shader 2 (edge smoothing) processing time by Kernel size.}
\label{shader2_execution_graph}
\end{figure}

Fig. \ref{shader2_execution_graph} shows the execution times of the edge smoothing algorithm for various kernel sizes.

%\begin{figure}[!t]
%\centering
%\includegraphics[width=3.2in]{fig24}
%\caption{Shader 2 (edge smoothing) typical execution time on GPU.}
%\label{shader2_execution}
%\end{figure}  
	
\section{Discussions}

\subsection{GPU Performance}
To execute complicated image improvement algorithms in real time, the LiDAR depth data integration with GPU processing requires balancing the computational load. LiDAR sensors provide depth data that can vary across an image. Efficient processing of this data is crucial, to maintain frame rates and ensure smooth operation. The edge smoothing feature is the most performance-intensive component of our implementation.
 
\subsection{The Image Signal Processor (ISP)}
The ISP in iPhones is designed and developed to perform a range of specialized tasks necessary for capturing and processing photos and videos from the camera's sensors. It performs essential functions that produce cleaner images, which include noise reduction, auto exposure, autofocus, and color correction.  
 
Image correction occurs automatically on the ISP before the image data reaches the GPU, and features like autofocus and auto exposure based on detected faces can be enabled in Swift to run on the ISP. We enabled both face-driven autofocus and auto exposure in our application, while also providing sliders on the user interface for further color channel and exposure adjustment by the user to allow for correction of both the background and foreground images (Fig. \ref{face_detection}). This allows the user to better match the foreground subject to the background image.

\subsection{Algorithmic Efficiency}
\begin{figure}[!t]
\centering
\includegraphics[width=3.2in]{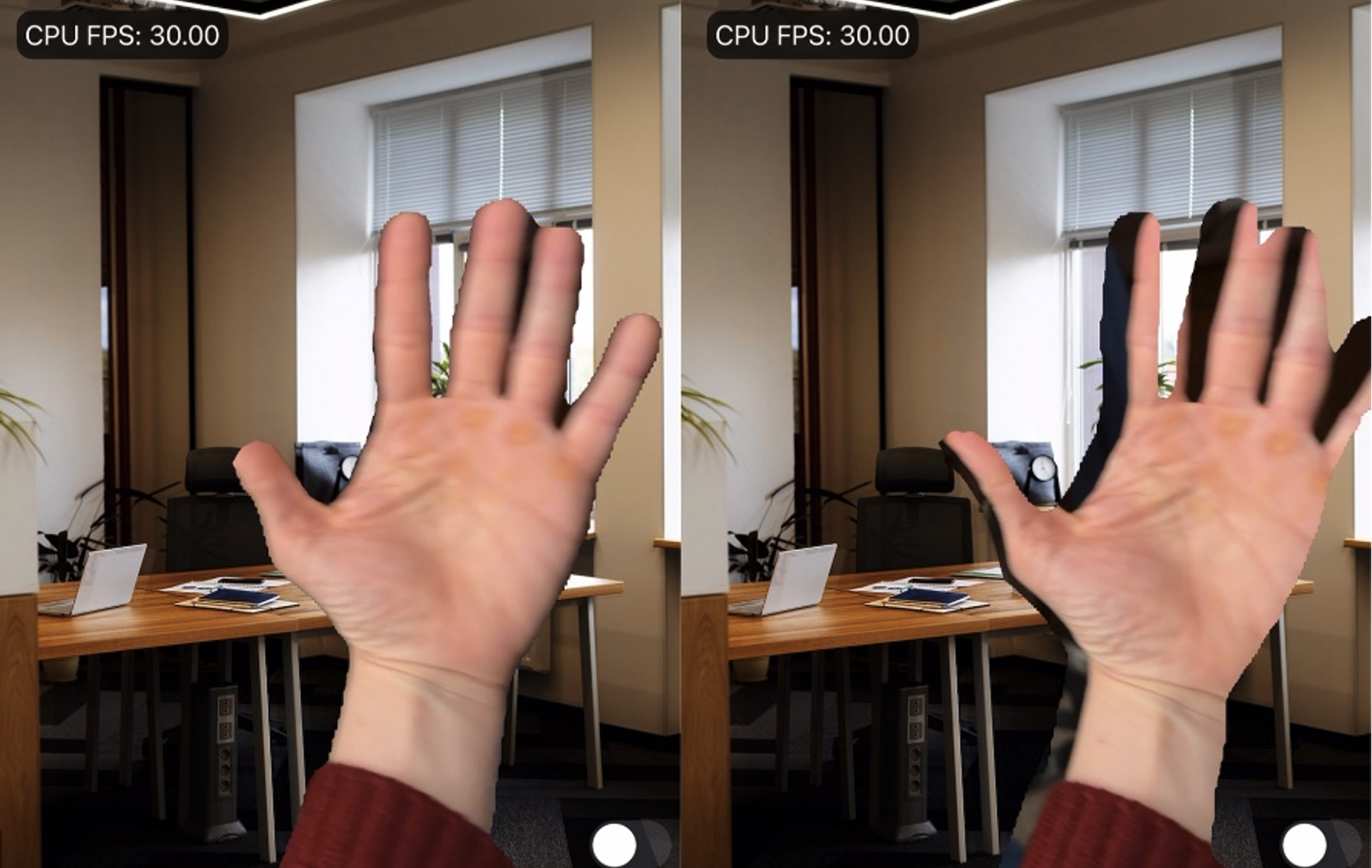}
\caption{Frame where color image is aligned with LiDAR image (left) and frame where updated 60fps color image is not aligned with reused 30fps LiDAR image (right).}
\label{liDAR_lag}
\end{figure}

Our approach has demonstrated the effectiveness and efficiency of combining cutting-edge GPU-based image processing techniques with LiDAR to improve photo and video quality on a live stream. The background removal and alpha roll-off process that takes place in Shader 1 takes less than 1ms, or less than 6\% of the GPU’s 16.67ms of allotted time for a 60fps screen refresh rate. Even Shader 2, when using a 9x9 kernel in the morphological close can still execute in under 7ms, or 41\% of the frame period, keeping the entire render sequence within the 16.67ms window. It is important to note that even though the color camera frame rate and refresh rate of the screen is 60fps, the LiDAR camera is bandwidth-limited to 30fps as shown in Fig. \ref{lidar_res_table}. This results in the GPU processing the same LiDAR data twice before a new frame is available. This lag in LiDAR data becomes visible in every other displayed composite image when a foreground subject moves quickly across the frame as shown in Fig. \ref{liDAR_lag}. This effect is not noticeable in stationary subjects. 

The ability to isolate subjects from backgrounds in video streams without requiring a green screen or controlled lighting represents a significant advancement. This approach focuses on features that ISPs do not cover, such as precise subject isolation in low-light.
The feature of integrating LiDAR's depth data for precise subject isolation is helpful, especially in public places where a subject may not want to include some or all of their background. Our app provides flexibility on the depth and roll-off the user may need for background removal or inclusion. It allows for selective visual data processing, minimizing extraneous data captured and simultaneously reducing the risk of capturing someone's private information without consent. It anonymizes individuals in video and image content, ensuring personal identities are protected in public or semi-public settings. Thus, this approach maintains privacy while accomplishing the desired photographic outcome.

\section{Conclusions and Future Works}
Our findings illustrate the effectiveness of the iPhone’s consumer-grade LiDAR in subject background removal and replacement, and the relative efficiency of performing this process on the GPU when compared to more computationally intensive processes like the morphological close operation used in the edge smoothing of the subject’s alpha channel. The LiDAR enables depth-based alpha roll-off, a novel feature that is not possible with traditional background removal techniques like chroma keying. The algorithms’ successful implementation and integration with existing mobile technology and frameworks have shown promising results for real-time background removal and replacement, particularly as an alternative to existing technologies where light levels cannot be controlled. Technological advances in GPU, CPU, and ISP performance that allow for higher LiDAR streaming resolutions will improve results and lessen the need for more computationally expensive image processing after the background removal step. For future work, this study’s image processing pipeline could be integrated into consumer video capture and live streaming apps, as well as augmented reality apps on any iPhones, iPads, or mobile devices leveraging integrated color and LiDAR cameras.

\section{Appendix}
All code for the iOS project can be found at \href{https://github.com/jkinnevan/Background-Removal-with-iPhone-15-LiDAR}{https://github.com/jkinnevan/Background-Removal-with-iPhone-15-LiDAR}

\bibliographystyle{IEEEtran}
\bibliography{references}

\end{document}